\documentclass[usenatbib,useAMSm a4paper, singlespacing]{mn2e}

\usepackage{subcaption}
\usepackage{caption}
\usepackage{graphicx}	% Including figure files
\usepackage{amsmath}	% Advanced maths commands
\usepackage{amssymb}	% Extra maths symbols
\usepackage{bm}		% Bold maths symbols, including upright Greek
\usepackage{pdflscape}
\usepackage[T1]{fontenc}
\usepackage{txfonts}
\usepackage{float}

\usepackage{longtable,lscape, booktabs,etoolbox}
\usepackage{rotating}
\usepackage{multirow,multicol}

\usepackage{epsfig}
\usepackage{footnote}
\usepackage[referable]{threeparttablex}
\usepackage[normalem]{ulem}
\usepackage{color}
\usepackage{url}
\usepackage[breaklinks]{hyperref}
\usepackage{dblfloatfix}
\usepackage{placeins}
\def\mearth{\ifmmode {\rm M_{\oplus}}\else $\rm M_{\oplus}$\fi}
\def\Mearth{\ifmmode {\rm M_{\oplus}}\else $\rm M_{\oplus}$\fi}
\def\Rearth{\ifmmode {\rm R_{\oplus}}\else $\rm R_{\oplus}$\fi}
\def\Ms{\ifmmode {M_s}\else $M_s$\fi}
\def\Mp{\ifmmode {M_p}\else $M_p$\fi}
\def\Rp{\ifmmode {R_p}\else $R_p$\fi}
\def\rearth{\ifmmode {\rm R_{\oplus}}\else $\rm R_{\oplus}$\fi}

\newcommand{\Msun}{M_{\odot}}

\newcommand{\Lsun}{L_{\odot}}

\newcommand{\onethird}{{\frac{1}{3}}}
\newcommand{\onehalf}{{\frac{1}{2}}}
\newcommand{\twothirds}{{\frac{2}{3}}}

\title[THE FREQUENCY OF BINARY STARS INTERLOPERS AMONGST TRANSITIONAL DISKS.]
{THE FREQUENCY OF BINARY STAR INTERLOPERS AMONGST TRANSITIONAL DISKS.}

\author[D. Ru\'iz-Rodr\'iguez et al.]{\Large D. Ru\'iz-Rodr\'iguez$^{1}$, M. Ireland$^1$,  L. Cieza$^{2,3}$,  A. Kraus$^{4}$\\
$^{1}$Research School of Astronomy and Astrophysics, Australian National University, Canberra, ACT 2611, Australia; dary.ruiz@anu.edu.au\\
$^{2}$N\'ucleo de Astronom\'ia, Facultad de Ingener\'ia, Universidad Diego Portales, Av. Ejercito 441, Santiago, Chile \\
$^{3}$Millenium Nucleus ``Protoplanetary Disks in ALMA Early Science", Chile\\
$^{4}$Department of Astronomy, The University of Texas at Austin, Austin, TX 78712, USA\\
}
\begin{document}

\date{}

\pagerange{\pageref{firstpage}--\pageref{lastpage}} \pubyear{2016}

\maketitle

\label{firstpage}

\begin{abstract}

% $\gtrsim$

Using Non-Redundant Mask interferometry  (NRM), we searched for binary companions to objects previously classified as Transitional Disks (TD). These objects are thought to be an evolutionary stage between an optically thick disk and optically thin disk. We investigate the presence of a stellar companion as a possible mechanism of material depletion in the inner region of these disks, which would rule out an ongoing planetary formation process in distances comparable to the binary separation. For our detection limits, we implement a new method of completeness correction using a combination of randomly sampled binary orbits and Bayesian inference. The selected sample of 24 TDs  belong to the nearby and young star forming regions: Ophiuchus ($\sim$ 130 pc),  Taurus-Auriga ($\sim$ 140 pc) and IC348 ( $\sim$ 220 pc). These regions are suitable to resolve faint stellar companions with moderate to high confidence levels at distances as low as 2 au from the central star. With a total of 31 objects, including 11 known TDs and circumbinary disks from the literature, we have found that a fraction of 0.38 $\pm$ 0.09 of the SEDs of these objects are likely due to the tidal interaction between a close binary and its disk,  while the remaining SEDs are likely the result of other internal processes such as photoevaporation, grain growth, planet disk interactions. In addition, we detected four companions orbiting outside the area of the truncation radii and we propose that the IR excesses of these systems are due to a disk orbiting a secondary companion

\end{abstract}


\section{Introduction}
\label{sec: intro}

After the formation of a star, the lifetime of a disk is estimated to be $\la$ 10 Myrs. At an age of $\sim$5 Myrs, around 90$\%$ of these objects already went through an evolution process of dispersion of their optically thick primordial disks  \citep{SiciliaAguilar2006}. The dispersion of the inner disk material creates unique morphologies in the disk that can be detected by their unusual spectral energy distributions (SED) \citep{Strom1989}. Assuming that all disks go through this dispersing phase, then approximately 10$-$20$\%$ of the disks are in a ``transition" phase with time-scales within $<$ 0.5 Myr;  \citep{Furlan2011, Koepferl2013}. In comparison with the characteristic continuum level of the SED of a Classical T Tauri Star (CTTS), these objects are defined as: stellar objects with small near-infrared (NIR) and/or mid-infrared (MIR) excesses and large MIR and/or far-infrared (FIR) excesses \citep[e.g.][]{Espaillat2014}. Given the ambiguity in the literature as to whether a disk in a ``transition phase" makes reference exclusively to a disk with an inner hole surrounding a single star or also includes binary systems in a transition phase, we will describe disks around single stars exclusively as \textit{Transitional Disk} (TD) and to describe disks around binary stars as \textit{Circumbinary Disks} (CD).

Detailed modelling of TD disk SEDs has interpreted the reduction of excess in the NIR-MIR as the dearth of small dust grains and thin gas in the inner region of the disk \citep{Espaillat2012}. In addition, mm-interferometric observations have mapped this particular disk morphology of the TDs, showing a dust-depleted region in the inner disk and/or gaps \citep{Andrews2011, Canovas2016}. Although the physical origins causing these particular shapes in the disks are still unclear, several theories have been proposed to explain the clearing mechanisms in the disk from inside out, such as grain growth \citep{Dullemond2001}, magnetorotational instability \citep{ChiangMurray2007}, photoevaporation \citep{Clarke2001, Alexander2007}, dust filtration \citep{Rice2006a}, and disk-planet(s) interactions \citep{Kraus2012, Dodson2011}. However, it has been difficult to reconcile the main process of dispersion of the disk, especially since these mechanisms might dominate at different time-scales and radii.
For instance, planet formation and photoevaporation may play a sequential dominating role in the disk dispersion phase, since photoevaporation disperses more rapidly once a planet is formed and has carved a gap in the disk \citep{Rosotti2015}. 

Unfortunately, these models are still not able to simultaneously explain the evolution process of all TDs, especially those with high accretion rates and large inner cavities full of large amounts of gas near the central star. However, fully understanding  the disk dispersal process is of a vital importance, because it provides insights about the formation of planetary systems like our own \citep{Dodson2011}. In particular, knowledge of the timescales of gas survival sets constraints on the time available for the formation of a gas rich planet via core accretion \citep{Pollack1996}. Alternatively, another clearing mechanism has been proposed for the truncation of the inner disk: the presence of a \textit{stellar companion}. \citet{Artymowicz1994} showed that in the binary$-$disk interaction, the stellar companion will truncate the CD at a distance, which depends highly on the eccentricity and mass ratio of the binary system. These theoretical models predict that the ratio of the inner radii ($r_{d}$) about the center of mass and the semi-major axis ($a$) of the binary system ranges from 1.7 to 3.3 for nearly circular orbits (e $=$ 0$-$0.25) and highly eccentric binaries ($e$ $\sim$ 0.75), respectively. Although, previous surveys of stellar companions in a range of $\sim$ 3 $-$ 50 au have indicated that binary truncation might not be a primary mechanism for the clearing inner region of the disk  \citep{Pott2010, Kraus2012}, there are different factors that prevented the detection of faint stellar companions in general, such as inner working angle and a small separation of the binary at the observing epoch. 

In addition, a misleading interpretation of the SEDs can occur in the classification process of TDs through the SEDs of the CDs. Since an unresolved faint infrared companion can aggregate NIR flux to the net SED and if this object is surrounded by a disk, it could emit MIR levels similar to the MIR excess seen in the SED of TDs \citep[e.g][]{Duchene2003, Kraus2015_FWTau}. Although, the SED of these CDs present several overlapping features with a ``normal" SED of TDs, it would be misleading to treat them in a similar way. For instance, the implications for the presence of another star in the star-disk system entails an incorrect measurement of the luminosity and temperature, which translates into inaccurate age and mass estimates. This is the case of Coku Tau/4  and CS Cha that were originally described as TDs \citep{Forrest2004, Espaillat2007}, but eventually were presented as CDs \citep{Ireland2008, Guenther2007}. This misclassification would be reflected in the estimation of birthplaces and timescales for formation of sub-stellar companions (brown dwarfs) and/or planetary systems, and the demographic properties of these populations \citep[e.g.][]{Najita2015}.

Therefore, determining a more accurate relative picture of the lifetime of TDs and CDs requires a comprehensive survey capable of resolving close binaries ($\la$ 30 au) and measuring their frequency in objects previously classified as TDs through their SEDs. Although, the open gap in the inner region of the disk might have different physical origins, in this paper we seek to identify if the dispersion of the primordial material in the inner region of the disk is a result of the tidal interaction between a close binary system and the disk. At small separations, detecting faint companions orbiting bright stars, that in addition, are surrounded by dusty material, can be challenging due to the high contrast between the companion and the primary star.  However, observations of objects at early ages provide favorable IR contrast ratios for the detection of so far, unresolved faint companions because of their intrinsically higher luminosity ($\Delta$K $<$ 5 mag). 

We use the \textit{Non-Redundant Mask interferometry} (NRM) technique and NIRC2 instrument located at the Keck II telescope, which offers a solution to reach angular resolutions with the necessary contrast and is resistant to speckle noise in the image by measuring a self-calibrating quantity known as \textit{closure-phase} \citep[e.g.][]{Martinache2011}. In order to achieve a higher accuracy in the detection limits of our data, the NRM completeness as a function of position and contrast utilizes a combination of a \textit{MonteCarlo Integration} approach, giving a randomly sample of artificial binary stars, and \textit{Bayesian Inference}, which uses prior probability density functions of the binary orbital parameters. We have restricted the selection of objects to regions with an age of $\sim$ 1 $-$ 3 Myrs and within a distance of about 220 pc. Taurus-Aurigae , IC348 region (Perseus) and Ophiucus star forming regions satisfy these criteria \citep{Loinard2008, Wilking2008}.
 
This article is arranged as follows. In Section 2. we present the motivation for the sample selection, description of observations together with the data analysis and a review of the target properties such as distance to the star-forming regions and estimations of the inner radii. A simple \textit{Bayesian} modelling analysis of these data is conducted in Section 3, with an emphasis on prior probabilities and description of binary and single models. The results of fitting to closure$-$phases in the $\chi^{2}$ minimization are synthesized with other information in the literature in Section 4. To perform a statistical Bayesian analysis of the fraction of the binarity as the main responsible mechanism opening the gaps in the TDs, we present a \textit{Jefreys Prior} and its posterior probability in Section 5. Based on that analysis and observational results, we attempt to reconcile the observations with theoretical predictions from tidal interaction models and possible scenarios of planetary formation in Section 6. Finally, we provide an overall review of the work done and results in Section 7.

\section{TARGET SELECTION, OBSERVATIONS AND PROPERTIES}
\label{Sec:Target}

In the past decade, the identified populations of T Tauri stars in the Taurus$-$Auriga, IC348 and Ophiuchus star forming regions have been well studied since the \textit{Spitzer} data enriched the knowledge of dust distributions in the disks, providing large samples of Young Stellar Object members. 
Thus, we selected a sample in terms of their decreased flux (with respect to the CTTS median) in the wavelength range between $\sim$ 3 and 24 $\mu$m, which tracks dust out to separations of at least $\sim$ 30 au. Our targets were selected based on clear inner regions in the disk seen in their SEDs. Sources with excess at wavelength in the range of $\sim$ 8 to 24 $\mu$m and a lack of excess between $\sim$ 3 to 5  $\mu$m are taken as disks with small or no dust excess in the inner region. Also, we included sources with strong emission between $\sim$ 3 to 5  $\mu$m, but with a small excess emission at $\sim$ 8 to 24 $\mu$m and excess beyond $\sim$ 40 $\mu$m. %, see Figure \ref{fig:images2sed}. 
The sample of objects were selected mostly from the work of \citet{Muzerolle2010, Cieza2010, Cieza2012, Espaillat2012}  and \citet{Rebollido2015}. These programs aimed to characterize SEDs for those objects in a ``transition'' phase and provided disk masses and accretion rates of the targets. The Two Micron All Sky Survey (2MASS) catalog Ks magnitudes are used to assign apparent magnitudes to these objects classified as TDs. To maximise sensitivity in our observations and achieve an image resolution of $\sim$ 20 mas, we make use of the Kp filter to probe binary separations in this limit \citep[e.g.][]{Kraus2012}.  Considering that the maximum Kp-Ks colour of our objects is only 0.10 mag \footnote{We use the relation $(\mathrm{K' - K}) \cong 0.11(\mathrm{H - K})$ from \citet{Vacca2007}. }, we did not apply a magnitude conversion because it is less than the combination of our uncertainties in contrast ratio and the effects of stellar variability. Our final target list shown in Table \ref{table:Properties} is composed of 24 transitional disks with R magnitudes brighter than 18 and  spectral types in the range of G3$-$M5. For this final target list of TDs, only 2MASS J04210934+2750368 and EM* SR 24S have known stellar companions at 770 mas ($\sim$ 108 au) \citep{Cieza2010} and at 6000 mas ($\sim$ 650 au) \citep{Simon1995}, respectively. These are not a close companion located in the range of our area of detection and do not affect the main purpose of the observations.

\begin{figure}
 \centering
  \includegraphics[width=0.52\textwidth]{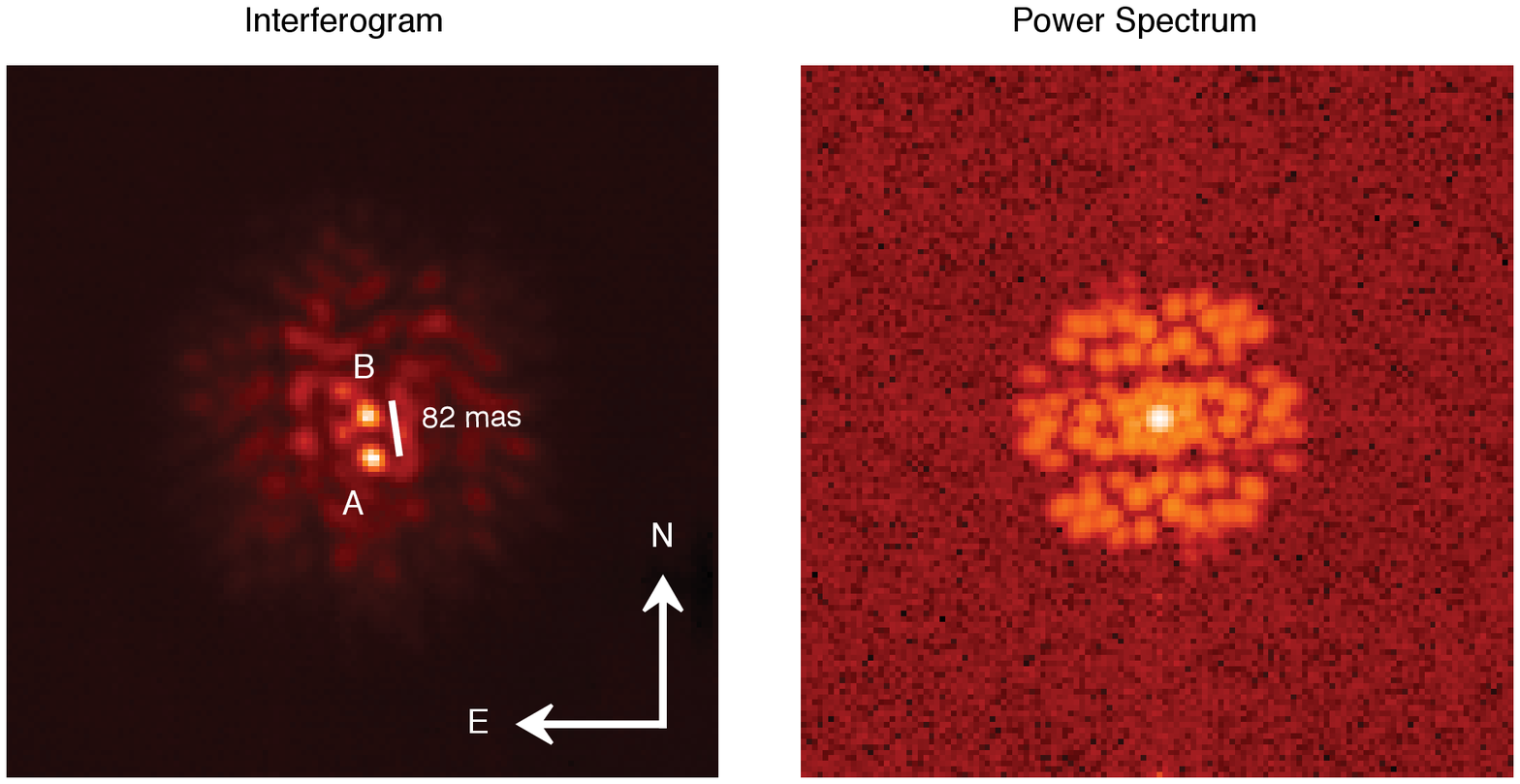}
   \includegraphics[width=0.48\textwidth]{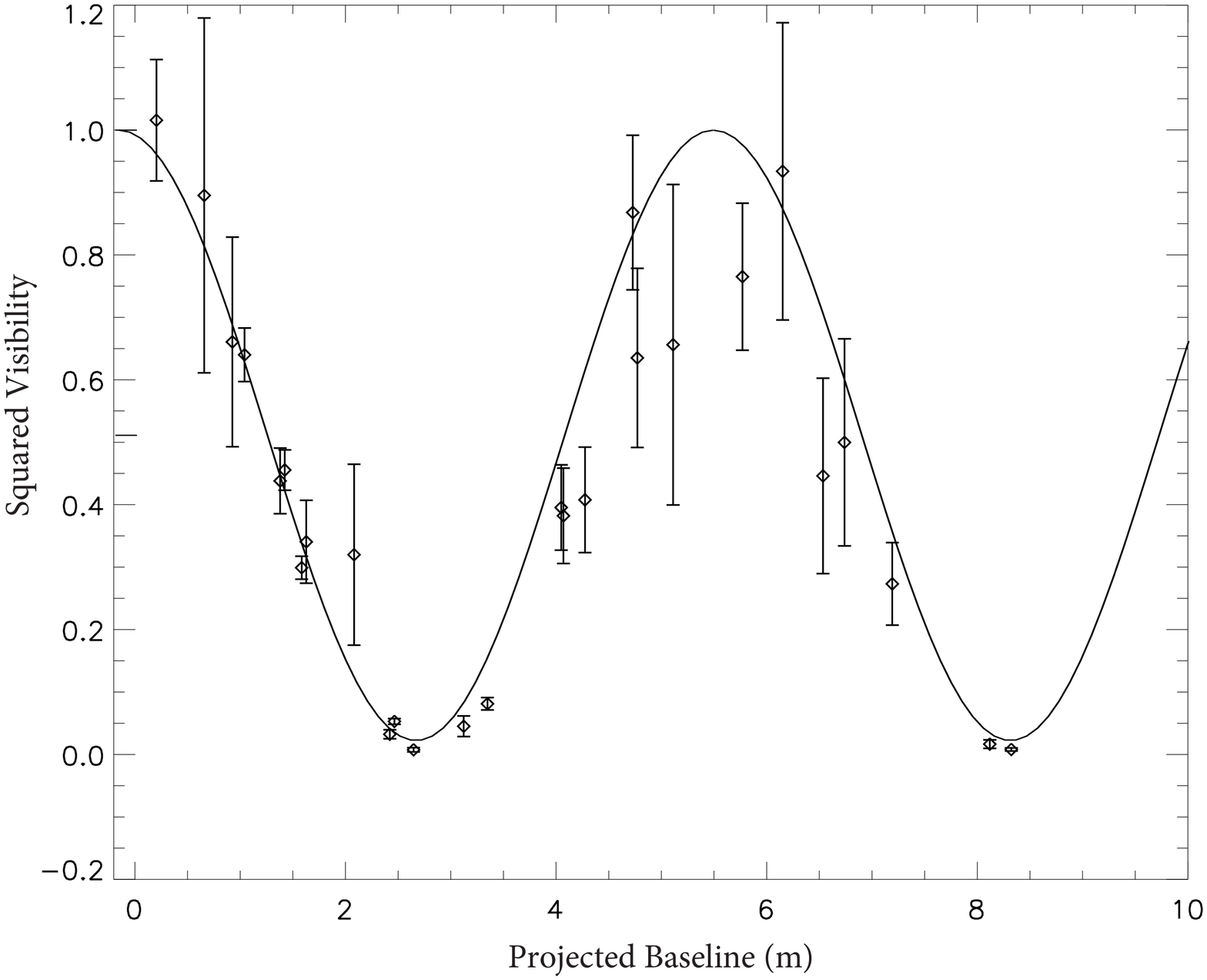}         
   \caption{Top: Interferogram and Power spectrum of the new reported binary LRL 135. Bottom: Squared visibilities as a function of the projected baseline. The solid line shows the best-fit of the binary parameters, angular separation and position angle.}
  \label{fig:interferogram}
\end{figure}

We observed our target list in August and December 2014 with the Adaptive Optics (AO) system of the near-infrared instrument (NIRC2) located at the Keck II 10 m telescope. The AO rotator tracking mode was set in vertical angle mode. A nine-hole mask located at the telescope pupil re-samples the light into a non-redundant interferogram of 36 pairwise fringes in the a Kp filter (Figure ~\ref{fig:interferogram}, top left panel). This pattern is specially designed to reach a near complete Fourier coverage. The Aladdin detector was configured to a 512 x 512 subarray and a multiple correlated double sampling readout mode was used in a narrow camera with a pixel scale of 9.952 $\pm$ 0.002 mas/pixel \citep{Yelda2010}. An overall exposure time of 20 seconds is used, except  for the calibrators LRL 410, CIDA 2 and UX Tau A with 60, 5 and 5 seconds, respectively. Because some TDs have been previously observed using identical settings as our observational method, we access the Keck Observatory Archive (KOA) and complete our sample of TDs that were unfit to observe in our 2014 runs. Table ~\ref{Table:observations} shows a summary of the observational settings of the targets, calibrators and their observing epochs.

The observed data have been corrected by flat-fielding, removal of bad pixels and dark subtracted to be spatially filtered with a super-Gaussian function to maximise sensitivity \citep{Ireland2008a}. The aperture masking analysis is based on the extraction and calibration of closure-phase and squared-visibility, then carrying out least squares binary fitting. The interested reader can find a detailed description available in e.g. \citet{Kraus2016}. In the case of fitting to binaries with an angular separation ($\rho$) of $\gtrsim$ 25 mas at high contrast, we fit only to closure-phase because of its immunity to changes in the AO point-spread function (PSF). We determined that any solutions with a significance of more than 6-$\sigma$ are detections of secondary components \citep{Kraus2016}. Then, we conducted a Bayesian analysis for marginal detections and contrast limits as described in section \ref{sec:bayes}. As an input to this Bayesian analysis, for each set of calibrated closure phases, we computed a least squares fit to contrast (secondary/primary flux) in a grid of 80 x 80 models with 5 milli-arcsec spacing. 

For reasons of both speed and in order to only consider the regime with symmetrical error bars, we approximated the contrast as being in the linear regime where closure phase is proportional to contrast. The output of this process was a grid of best fit contrasts and uncertainties. The uncertainties were scaled in order to acquire a $\chi^{2}$ equal to unity in the fit to the closure phases. For detections with $\rho$ $<$ 40 mas and $\Delta$K $<$ 1 mag, we included visibility amplitudes for breaking contrast/separation degeneracies. When fitting to squared visibilities, we conservatively added a 20$\%$ miscalibration uncertainty in quadrature to the uncertainties estimated from scatter in our data, and also left, as a free parameter, the scaling of the interferometric visibilities.  This was necessary because in AO data, Strehl ratios typically vary from target to PSF calibrator, and visibility amplitudes calibrate much more poorly than closure-phases. This miscalibration uncertainty needed to be added because Strehl variations, caused by e.g. changing atmospheric conditions, cause a variation in the visibility amplitudes between target and calibrator observations. Part of this uncertainty was taken into account by adding the scatter amongst calibrators in quadrature to the visibility amplitude uncertainty from the target.

Additionally, the detection limits are highly dependent on the contemporaneous observations of \textit{calibrators} that must be single stars with high S/N and ideally close to the observed target. The calibrators are used to remove effects of optical aberrations. Raw object visibilities are divided by calibrator visibilities, and calibrator closure phases are subtracted from raw object closure phases. However, we were not able to observe truly isolated stars in these dusty star-forming regions and for those observing epochs taken from KOA, we used likely single stars with non-redundant interferograms taken in the same observing run (Table ~\ref{Table:observations}). Therefore, in order to assure high S/N and non$-$binarity in the set of objects to be used as calibrators in each epoch, we perform the following steps:

\onecolumn
\begin{landscape}
\renewcommand{\thefootnote}{\fnsymbol{footnote}}
\renewcommand{\arraystretch}{0.84}
\begin{longtable}{lccccccccccl}
\caption[General Properties of Transitional Disks]{General Properties of Transitional Disks }
\label{table:Properties} \\
\hline \hline \\[-2ex]
   \multicolumn{1}{c}{\textbf{Target}} &
    \multicolumn{1}{c}{\textbf{Alter. Name}} &
   \multicolumn{1}{c}{\textbf{R.A. (J2000)}} &
   \multicolumn{1}{c}{\textbf{Dec (J2000)}} &
   \multicolumn{1}{c}{\textbf{R }}  &
   \multicolumn{1}{c}{\textbf{J }} &
   \multicolumn{1}{c}{\textbf{H }} &
   \multicolumn{1}{c}{\textbf{Ks  }} &
   \multicolumn{1}{c}{\textbf{Sp. Type}} &
   \multicolumn{1}{c}{\textbf{Log(Acc. Rate)}} &
   \multicolumn{1}{c}{\textbf{Inner Radii}\footnotemark[2]} &
   \multicolumn{1}{c}{\textbf{Reference}\footnotemark[1]} \\[0.8ex] 
     \multicolumn{1}{c}{\textbf{}} &
    \multicolumn{1}{c}{\textbf{}} &
   \multicolumn{1}{c}{\textbf{[$^{\mathbf{h}}$    $^{\mathbf{m}}$    $^{\mathbf{s}}$]}} &
   \multicolumn{1}{c}{\textbf{[$^{\circ}$ $ {}'$ ${}''$]}} &
   \multicolumn{1}{c}{\textbf{[mag]}} &
   \multicolumn{1}{c}{\textbf{[mag]}} &
   \multicolumn{1}{c}{$\mathbf{[mag]}$} &
   \multicolumn{1}{c}{$\mathbf{ [mag]}$} &
   \multicolumn{1}{c}{\textbf{}} &
   \multicolumn{1}{c}{\textbf{$\mathbf{ [ \,\Msun yr^{-1} ]}$}} &
   \multicolumn{1}{c}{\textbf{[au]}} &
   \multicolumn{1}{c}{\textbf{}} \\[1.5ex]\hline \hline \\[-2ex]
\endfirsthead

\endhead

\endfoot

\hline\hline\\[0.9ex] 
\endlastfoot

\\[0.2ex] 

\multicolumn{12}{c}{IC 348 }\\
\\[-0.2ex] \hline \\[-1.0ex]
LRL 21\footnotemark[4] &...& 03 44 56.15&+32 09 15.50 &14.81&11.02&9.99&9.47&K0&-9.4& 9 &1,2,3,4,16\\
LRL 67 &...& 03 43 44.62 &+32 08 17.90  &14.65& 12.05 & 11.13& 10.79&M0.75&-10.2& 10 &1,2,3,4,16\\
LRL 72&...&03 44 22.57 & +32 01 53.70  &15.93& 12.12 &  11.15&10.79&M2.5&$<$-11& 5 &1,2,3,4,16\\
LRL 237&...&03 44 23.57&+32 09 34.00 &17.72&13.50 & 12.74&12.40&M5&n& 0.005 &1,2,3,5,15\\
LRL 97&...&03 44 25.56&+32 06 17.00 &18.32&12.98 & 11.70 &11.14&M2.25&n& 0.005 &1,2,3,5,15\\
LRL 31&...&03 44 18.17&+32 04 57.00&17.22&12.09 & 10.54& 9.69&G6 &-7.9& 14 &1,2,3,4,6,16\\
LRL 182&...&03 44 18.20&+32 09 59.30&18.10&13.22 & 12.27&11.87&M4.25&n& -- &1,2,3,5\\
LRL 213&...&03 44 21.27&+32 12 37.30&16.78&13.70 & 12.92&12.51&M4.75&n& -- &1,2,3,5\\
LRL 58 &...&03 44 38.55&+32 08 00.70&16.43&11.94 & 10.90&10.47&M1.25&n& -- &1,2,3,5\\
LRL 135&...& 03 44 39.19& +32 20 09.00&16.90&12.65 & 11.80&11.44&M4.5&y& -- &1,2,3,5\\[0.2ex]\hline\\[0.2ex]

\multicolumn{12}{c}{Taurus-Aurigae }\\
\\[0.2ex]\hline\\[-1.0ex]

IRAS04125+2902&...& 04 15 42.79	&+29 09 59.77&14.34&10.71& 9.76&9.38&M1.25&-9.5&18 $-$ 24 &1,2,7,8,12\\
V410 X-ray 6&[GBA2007] 527 &04 19 01.11&+28 19 42.05&16.50&10.53& 9.60& 9.13&M4.5&-10.85& -- &1,2,9, 12\\
J04210934+2750368\footnotemark[5]&...&04 21 09.34&+27 50 36.84&15.70&11.23 & 10.66&10.36&M5.25&-10.3& -- &1,2,9\\[0.2ex]\hline\\[0.2ex]

\multicolumn{12}{c}{Ophiuchus }\\
\\[0.2ex]\hline\\[-1.5ex]

EM* SR 24S\footnotemark[3]&... & 16 26 58.51 & -24 45 36.87 & 14.15 & 9.75 & 8.17 & 7.06 & K2 &  -8.0 & 29 &1,17  \\ 
EM* SR 21A &...& 16 27 10.28 & -24 19 12.74 & 13.50 & 8.74 &  7.51& 6.72 & G3 & $<$ -9.0& 25  & 1,2, 13 \\
WSB 12&... &16 22 18.52&-23 21 48.10&13.03& 9.52&  8.59& 8.11&K5&-8.0& -- &1,2,10\\
 J16262367-2443138&DoAr 25 &16 26 23.68&-24 43 13.86&12.99&9.40& 8.40& 7.85&K5&-7.2& -- &1,2,10\\ 
 J16273901-2358187&DoAr 33 &16 27 39.01&-23 58 18.70&13.88&9.90 &8.72 &8.21&K5.5& -9.6& -- &1,2,10\\ 
WSB 63&... &16 28 54.07& -24 47 44.20&15.41& 10.68& 9.43 &8.91&M1.5&-8.1& 1.9 $\pm$ 0.3 &1,2,10, 14\\ 
 J16335560-2442049&RX J1633.9-2442&16 33 55.61&-24 42 05.00&15.04&10.46 &  9.36 &8.80&K7&-9.9& 7.9 $\pm$ 2.3&1,2,10,14\\ 
 J16250692-2350502&...&16 25 06.91 &-23 50 50.30&15.55&11.05 & 9.97 &9.51&M3&-8.8& 4.8 $\pm$ 2.5 &1,10,14\\
 J16315473-2503238&WSB 74 &16 31 54.73& -25 03 23.80&15.08& 10.14 & 8.66 &7.75&K7&-7.2& -- &1,10\\
WSB 40&...&16 26 48.65 &-23 56 34.20&15.42&10.43 &  9.18 & 8.45&K5.5&--& -- &1,2,11\\\
V* V852 Oph&...&16 25 24.34 &-24 29 44.30 & 14.52 & 10.75 &  9.87 & 9.45&M4.5&--& -- &1,2,11 
\footnotetext[1]{References: (1) 2MASS All-Sky Point Source Catalog, (2) \citet{Cutri2003}, (3) \citet{Luhman2003}, (4) \citet{Espaillat2012}, (5) \citet{Muzerolle2010}, (6) \citet{Flaherty2011}, (7) \citet{Luhman2009} , (8) \citet{Espaillat2015}, (9)  \citet{Cieza2012} , (10)  \citet{Cieza2010} , (11)  \citet{Rebollido2015},  (12)  \citet{Furlan2011}, (13) \citet{vanderMarel2016}, (14) \citet{Orellana2012}, (15) \citet{LeBlanc2011}, (16) \citet{Espaillat2010}, (17)  \citet{Andrews2011}}
\footnotetext[2]{Inner radii from literature.}
\footnotetext[3]{Stellar parameters taken from \citet{Andrews2011}.}
\footnotetext[4]{Target IDs of IC348 members are taken from the acronym Cl* IC 348 from \citet{Luhman1998}.}
\footnotetext[5]{Targets with 2MASS identifiers are presented by their designation e.g. J04210934+2750368.}

\end{longtable}
\end{landscape}

\clearpage
\newpage
\onecolumn
\renewcommand{\thefootnote}{\fnsymbol{footnote}}
\renewcommand{\arraystretch}{0.81}
\begin{longtable}{lccccccl}
\caption[Summary of Observations]{Summary of Observations.}
\label{Table:observations} \\
\hline \hline \\[-0.9ex]
   \multicolumn{1}{c}{\textbf{ID}} &
   \multicolumn{1}{c}{\textbf{BJD}} &
   \multicolumn{1}{c}{\textbf{t$_{int}$}} &
   \multicolumn{1}{c}{\textbf{Coadds}} &
   \multicolumn{1}{c}{\textbf{N$_{frames}$}}  &
   \multicolumn{1}{c}{\textbf{Airmass}} &
   \multicolumn{1}{c}{\textbf{Type}} &
   \multicolumn{1}{c}{\textbf{Note}} \\[0.5ex] 
    \multicolumn{1}{c}{\textbf{}} &
   \multicolumn{1}{c}{\textbf{(2400000 +)}} &
   \multicolumn{1}{c}{\textbf{[s]}} &
   \multicolumn{1}{c}{\textbf{}} &
   \multicolumn{1}{c}{$\mathbf{}$} &
   \multicolumn{1}{c}{$\mathbf{ }$} &
   \multicolumn{1}{c}{\textbf{}} &
   \multicolumn{1}{c}{\textbf{}} \\[1.5ex]\hline \hline \\[-2ex]
\endfirsthead

\multicolumn{7}{c}{{\tablename} \thetable{} -- Continued}\\[0.0ex]
  \hline \hline \\[-0.9ex]
   \multicolumn{1}{c}{\textbf{ID}} &
   \multicolumn{1}{c}{\textbf{BJD}} &
   \multicolumn{1}{c}{\textbf{t$_{int}$}} &
   \multicolumn{1}{c}{\textbf{Coadds}} &
   \multicolumn{1}{c}{\textbf{N$_{frames}$}}  &
   \multicolumn{1}{c}{\textbf{Airmass}} &
   \multicolumn{1}{c}{\textbf{Type}} &
   \multicolumn{1}{c}{\textbf{Note}} \\[0.5ex] 
   \multicolumn{1}{c}{\textbf{}} &
   \multicolumn{1}{c}{\textbf{(2400000 +)}} &
   \multicolumn{1}{c}{\textbf{[s]}} &
   \multicolumn{1}{c}{$\mathbf{}$} &
   \multicolumn{1}{c}{$\mathbf{ }$} &
   \multicolumn{1}{c}{\textbf{}} &
   \multicolumn{1}{c}{\textbf{}} \\[1.5ex]\hline \hline \\[-2ex]
   \endhead
\endfoot

\hline\hline\\[-0.5ex] 
\endlastfoot

\\[-0.5ex] 

\multicolumn{7}{c}{June 18, 2008 }\\
\\[-0.5ex]\hline

Haro 1-6  &     54635.75 &    5.00 &       4 &       7 &    1.92& Calibrator& $>$5$\sigma$\\
RX J1620.9-2352  &     54635.75 &    5.00 &       4 &       7 &    1.82 & Calibrator&\\
EM* SR 24S  &     54635.75 &    5.00 &       4 &       6 &    1.86 & Target& \\
EM* SR 24S  &     54635.75 &    5.00 &       4 &       8 &    1.81& Target& \\
Haro 1-6  &     54635.75 &    5.00 &       4 &       7 &    1.72 & Calibrator&\\
Haro 1-6  &     54635.79 &    1.00 &      10 &      11 &    1.53 & Calibrator&\\
EM* SR 24S  &     54635.79 &    1.00 &      10 &       7 &    1.51 & Target& \\
EM* SR 21A  &    54635.88 &    1.00 &      20 &       7 &    1.40 & Target& \\
EM* SR 24N &    54635.88 &    2.50 &       8 &       2 &    1.41 & Calibrator&$>$5$\sigma$\\
J16262367-2443138  &     54635.88 &    2.50 &       8 &       7 &    1.42 & Target& \\
WSB 12  &     54635.88 &    2.50 &       8 &       7 &    1.40 & Target& \\
EM* SR 21A  &     54635.88 &    2.50 &       8 &       7 &    1.43 & Target& \\
J16262367-2443138    &     54635.88 &    2.50 &       8 &       7 &    1.45 & Target& \\
WSB 12 &54635.88 &    2.50 &       8 &       7 &    1.43 & Target& \\
EM* SR 21A &     54635.92 &    2.50 &       8 &       7 &    1.47 & Target& \\
J16262367-2443138  &     54635.92 &    2.50 &       8 &       7 &    1.50  &Target& \\
WSB 12  &     54635.92 &    2.50 &       8 &       7 &    1.49 & Target& \\		
V* V2059 Oph  &    54635.92 &   20.00 &       1 &       7 &    1.52  & Calibrator&$>$5$\sigma$\\
RX J1625.2-2455  &    54635.92 &   20.00 &       1 &       7 &    1.58 & Calibrator&\\
V* V2059 Oph  &    54635.92 &   20.00 &       1 &       7 &    1.61 & Calibrator&\\[0.2ex]\hline\\[0.2ex]

\multicolumn{7}{c}{November 03, 2009 }\\
\\[0.2ex]\hline\\

J04183030+2743208  &    55138.92 &   10.00 &       1 &      10 &    1.05 & calibrator \\
J04380083+2558572  &    55138.96 &   10.00 &       1 &      10 &    1.01 & calibrator & \\
J04350850+2311398  &    55139.00 &   20.00 &       1 &      10 &    1.01& calibrator & \\
V410 X-ray 6 &                 55139.00 &   10.00 &       1 &       9 &    1.03  & Target\\
J04244506+2701447  &    55139.04 &   10.00 &       1 &       9 &    1.05  & calibrator \\[-0.1ex]\hline\\[-0.1ex]

\multicolumn{7}{c}{April 23, 2011 }\\
\\[-0.5ex]\hline\\

 J16233462-2308467   &     55674.92 &    5.00 &       4 &      12 &    1.67  & Calibrator& $>$5$\sigma$\\
 WSB 63 &     55674.92 &    5.00 &       4 &      10 &    1.65  & Target & \\ 
J16273901-2358187  &     55674.96 &    5.00 &       4 &       9 &    1.55 & Target & \\
J16250692-2350502   &     55674.96 &    5.00 &       4 &      10 &    1.49  & Target & \\
DoAr 32   &     55674.96 &    5.00 &       4 &      11 &    1.47 & Calibrator & $>$4$\sigma$\\
WSB 63  &     55674.96 &    5.00 &       4 &      10 &    1.46 & Target & \\  
BKLT J162624-244323  &     55674.96 &    5.00 &       4 &      11 &    1.44  & Calibrator & $>$5$\sigma$\\
 WSB12 (RX J1622.3-2321)  &     55675.00 &    5.00 &       4 &      11 &    1.38 & Target & \\
 J16315473-2503238   &     55675.00 &    5.00 &       4 &      10 &    1.42 & Target &\\
BKLT J162624-244323  &    55675.00 &    5.00 &       4 &      10 &    1.40 & Calibrator & \\
 EM* SR 8  &     55675.00 &    5.00 &       4 &      10 &    1.40 & Calibrator &\\
 J16335560-2442049  &    55675.00 &    5.00 &       4 &      11 &    1.40 & Target & \\
 DoAr 50  &     55675.00 &    5.00 &       4 &      10 &    1.42 & Calibrator &\\
 WSB 63 &    55675.04 &    5.00 &       4 &      10 &    1.41  & Target &  \\
 J16335560-2442049   &     55675.04 &    5.00 &       4 &      10 &    1.41  & Target & \\
  DoAr 50  &     55675.04 &    5.00 &       4 &       9 &    1.44 & Calibrator & $>$5$\sigma$\\
  DoAr 24  &    55675.04 &    5.00 &       4 &      10 &    1.44 & Calibrator &  \\
  2E 1624.2-2444 &    55675.04 &    5.00 &       4 &      11 &    1.49 & Calibrator & \\
  RX J1633.9-2442   &     55675.08 &    5.00 &       4 &      10 &    1.48  & Target & \\
  RX J1624.8-2359 &    55675.08 &    5.00 &       4 &      11 &    1.51 & Calibrator & no-Binary\\
  ROXs 4 &     55675.08 &    5.00 &       4 &      11 &    1.57 & Calibrator & \\
  2E 1624.6-2352  &    55675.08 &    5.00 &       4 &      11 &    1.60  & Calibrator & \\
  IRAS 16114-1858 &     55675.08 &    5.00 &       4 &       9 &    1.62 & Calibrator & \\[-0.1ex]\hline\\[-0.1ex]

\multicolumn{7}{c}{November 15, 2011}\\
\\[-0.5ex]\hline\\

 MBO 22  &     55880.75 &    5.00 &       4 &      25 &    1.50 & Calibrator&$>$5$\sigma$\\
LRL 21  &     55880.79 &    5.00 &       4 &      23 &    1.42 & Target\\
LRL 72  &     55880.79 &    5.00 &       4 &      23 &    1.33 & Target & \\
MBO 22   &     55880.79 &    5.00 &       4 &      23 &    1.22 & Calibrator& \\
LRL 21  &     55880.83 &    5.00 &       4 &      23 &    1.22 & Target\\
LRL 67  &     55880.83 &    5.00 &       4 &      21 &    1.18 & Target\\
LRL 21  &     55880.83 &    5.00 &       4 &      22 &    1.11 & Target\\
MBO 22  &     55880.83 &    5.00 &       4 &       2 &    1.07  & Calibrator& \\
LRL 67  &     55880.88 &    5.00 &       4 &      23 &    1.06 & Target\\
J03302409+3114043  &     55880.88 &    5.00 &       4 &      23 &    1.03 & Calibrator& $>$5$\sigma$\\
V410 X-ray 6   &     55880.88 &    5.00 &       4 &      23 &    1.05 & Target&\\
J04300424+3522238 &     55880.92 &    5.00 &       4 &      18 &    1.07 & Calibrator&$>$5$\sigma$\\
V410 X-ray 6  &     55880.92 &    5.00 &       4 &      21 &    1.02  & Target\\
J04300424+3522238  &    55880.92 &    5.00 &       4 &      19 &    1.04  & Calibrator&\\
HBC 390 &     55880.96 &    5.00 &       4 &      22 &    1.04   & Calibrator&\\
J04303235+3536133   &     55880.96 &    5.00 &       4 &      10 &    1.04 & Calibrator&no-Binary\\
J04303235+3536133  &     55880.96 &    5.00 &       4 &      11 &    1.04 & Calibrator&no -Binary\\
J03302409+3114043  &     55880.96 &    5.00 &       4 &      14 &    1.09 & Calibrator&$>$5$\sigma$\\[-0.1ex]\hline\\[-0.1ex]

\multicolumn{7}{c}{April 14, 2012}\\
\\[-0.5ex]\hline\\

 RX J1615.3-3255   &     56032.04 &    5.00 &       4 &      12 &    1.66 & Calibrator&\\
 RX J1615.9-3241  &     56032.04 &    5.00 &       4 &      13 &    1.66 & Calibrator&\\
 RX J1625.3-2402 &     56032.08 &    5.00 &       4 &       3 &    1.42 & Calibrator&no-HighBF\\
 V* V852 Oph  &     56032.08 &    5.00 &       4 &       3 &    1.45 & Target\\
 WMR2005 1-38  &    56032.08 &    5.00 &       4 &       3 &    1.47  & Calibrator&\\
  WSB 40  &     56032.08 &    5.00 &       4 &       3 &    1.50 & Target &\\
 WMR2005 1-21  &     56032.12 &    5.00 &       4 &       3 &    1.61   & Calibrator&\\[-0.1ex]\hline\\[-0.1ex]

\multicolumn{7}{c}{August 11, 2014}\\
\\[-0.5ex]\hline\\

V* V711 Per &    56881.04 &   20.00 &       1 &       8 &    1.38 & Calibrator&$>$5$\sigma$\\
LRL 110   &    56881.04 &   20.00 &       1 &       6 &    1.35   & Calibrator&\\
LRL 410  &    56881.04 &   60.00 &       1 &       5 &    1.31   &Calibrator& \\
CIDA 2  &     56881.08 &    5.00 &       4 &       8 &    1.36    & Calibrator&$>$5$\sigma$\\
IRAS 04125+2902  &    56881.12 &   20.00 &       1 &       7 &    1.16  & Target &\\
V410 X-ray 3  &    56881.12 &   20.00 &       1 &       5 &    1.13   & Calibrator&$>$5$\sigma$\\
UX Tau A  &     56881.12 &    5.00 &       4 &       8 &    1.14   & Calibrator&\\[-0.1ex]\hline\\[-0.1ex]

\multicolumn{7}{c}{August 12, 2014}\\
\\[-0.5ex]\hline\\

LRL 75  &    56882.04 &   20.00 &       1 &       8 &    1.54   & Calibrator&\\
LRL 40  &    56882.04 &   20.00 &       1 &       8 &    1.46   & Calibrator&\\
 LRL 168  &    56882.04 &   20.00 &       1 &       8 &    1.42  & Calibrator&\\
 LRL 97   &    56882.04 &   20.00 &       1 &       8 &    1.35  & Target&\\
 LRL 237  &   56882.04 &   20.00 &       1 &       6 &    1.31 & Target&\\[-0.1ex]\hline\\[-0.1ex]

\multicolumn{7}{c}{August 13, 2014}\\
\\[-0.5ex]\hline\\

KOI-137  &    56883.00 &   20.00 &       1 &       7 &    1.65 & Calibrator&\\
KOI-0044  &   56883.00 &   20.00 &       1 &       6 &    1.66  & Calibrator&\\
KOI-4567  &    56883.00 &   20.00 &       1 &       7 &    1.66  & Calibrator&\\
LRL 72 &    56883.04 &   20.00 &       1 &       7 &    1.55 & Target&\\
J04311907+2335047  &    56883.08 &   20.00 &       1 &       7 &    1.37 & Calibrator& $>$5$\sigma$\\[-0.5ex]\hline\\[-0.5ex]

\multicolumn{7}{c}{December 09, 2014}\\
\\[-0.5ex]\hline\\

LRL 53   &    57000.75 &   20.00 &       1 &       7 &    1.19  & Calibrator&\\
LRL 182 &    57000.75 &   20.00 &       1 &       8 &    1.17 & Target& \\
LRL 58  &    57000.79 &   20.00 &       1 &       8 &    1.13 & Target&\\
LRL 355  &    57000.79 &   20.00 &       1 &       7 &    1.12 & Calibrator&\\
LRL 135  &    57000.79 &   20.00 &       1 &       7 &    1.10  & Target&\\
LRL 233  &    57000.79 &   20.00 &       1 &       7 &    1.08 & Calibrator& $>$5$\sigma$\\
LRL 31  &    57000.79 &   20.00 &       1 &       4 &    1.07  & Target&\\
LRL 169  &    57000.79 &   20.00 &       1 &       7 &    1.06   & Calibrator&\\
LRL 213  &   57000.83 &   20.00 &       1 &       7 &    1.05  & Target&\\
J03302409+3114043  &    57000.96 &   20.00 &       1 &      13 &    1.41 & Calibrator&$>$5$\sigma$\\
MBO 22   &   57001.00 &   20.00 &       1 &       7 &    1.46  & Calibrator& $>$5$\sigma$\\
 J04210934+2750368 &    57001.00 &   20.00 &       1 &       3 &    1.25 & Target&\\
J04300424+3522238  &    57001.00 &   20.00 &       1 &       8 &    1.30  & Calibrator&\\
J04303235+3536133  &    57001.00 &   20.00 &       1 &       6 &    1.33 & Calibrator&no-binary\\
Pr 0211  &    57001.08 &   20.00 &       1 &       7 &    1.02  & Calibrator&\\
Pr 0225  &    57001.08 &   20.00 &       1 &       7 &    1.02 & Calibrator&

\end{longtable}

\normalsize
\renewcommand{\thefootnote}{\arabic{footnote}}
\renewcommand{\arraystretch}{1.0}

\twocolumn

\begin{enumerate}
\item We first identified the set of targets of obvious binarity, e.g LRL 135 shown in Figure \ref{fig:interferogram}, and targets with significance levels of $>$ 8-$\sigma$ by fitting only to closure-phase and then, removed them from the source sample of calibrators.
\item The remaining objects, calibrators and science targets, play the role of \textit{inter-calibrating sources}. After fitting closure phases for every observed object, we started eliminating from the set of calibrator sources those objects with a significance of more than 5-$\sigma$, assuring an isolated object with high S/N.
\item The closure phases of those remaining objects are used as the final set of calibrators.
\end{enumerate}

In our survey, the aperture masking data identified a well resolved a nearly equal luminosity companion for LRL 135 at 82 mas with a $\Delta$K of 0.17 $\pm$ 0.01 and a position angle of 208$^{^{\circ}}$, which is shown in Figure ~\ref{fig:interferogram} as the observed interferogram and its power spectrum. We provide a more detailed description about these results in Section \ref{Sec:companionoutside}, including reference to maser distances. Also, the extracted squared visibilities are plotted, where clearly fringe contrast goes to near zero at the longest baseline of the mask ($\sim$ 9 m).

\subsection{Target Properties}
\label{sec:stellarparameter}

Our methodology to identify companions that might be responsible for the observed TD SEDs and could be orbiting in the inner region of the disk, requires estimations of the distances to the inner disk wall from the central star ($r_{d}$). However, not all TDs have a previous measurements of the inner radii and are not calculated by following a standard approach. Because we seek for uniformity in these estimations, we developed a simple approach highly dependent on the stellar luminosities ($L_{\star}$) and the dust temperature in the disk ($T_{d}$). The stellar luminosities are calculated with the dereddened $J$-band photometry method from \citet{Kenyon1995} and adopting the known distances to each different star$-$forming region. We de-reddened the $J$-band fluxes using the A$_{J}$  extinction and measured by following the \citet{Mathis1990} extinction law  with $A(\lambda )/A(J)\sim (\lambda /1.25)^{-\alpha}$, where $\alpha=1.7$. We used $A_{J}=2.62[(J-H)-(J-H)_{0}]$, where $(J-H)_{0}$ is the expected colour of a main-sequence star from \citet{Pecaut2013}. We derived the stellar properties based on the spectral types taken from literature and a conversion to the effective temperatures (T$_{\rm eff}$) taken from \citet{Pecaut2013} with uncertainties of  $\sim$ 150 K, corresponding to $\pm$ 1 spectral subclass (Table \ref{table:Properties}). Additionally, using T$_{\rm eff}$ and L$_{\star}$, and assuming a single star system, we estimated the stellar masses ($M_{\star}$) for each TD. Those in the range between 0.01 and 1.4 $M_{\odot}$, were derived from \citet{Baraffe2015} and stellar masses $>$1.4 $M_{\odot}$ from the PARSEC evolutionary models \citep{Bressan2012}. For the unknown metallicity values we adopted solar composition, and we also held the surface gravity fixed at the value log $\textrm{g}=$ 4.0, typical for PMS stars. Table \ref{table:Properties} shows R and Ks magnitudes, spectral types and stellar properties of these objects.

\begin{figure*}
  \centering
     \includegraphics[width=0.65\textwidth]{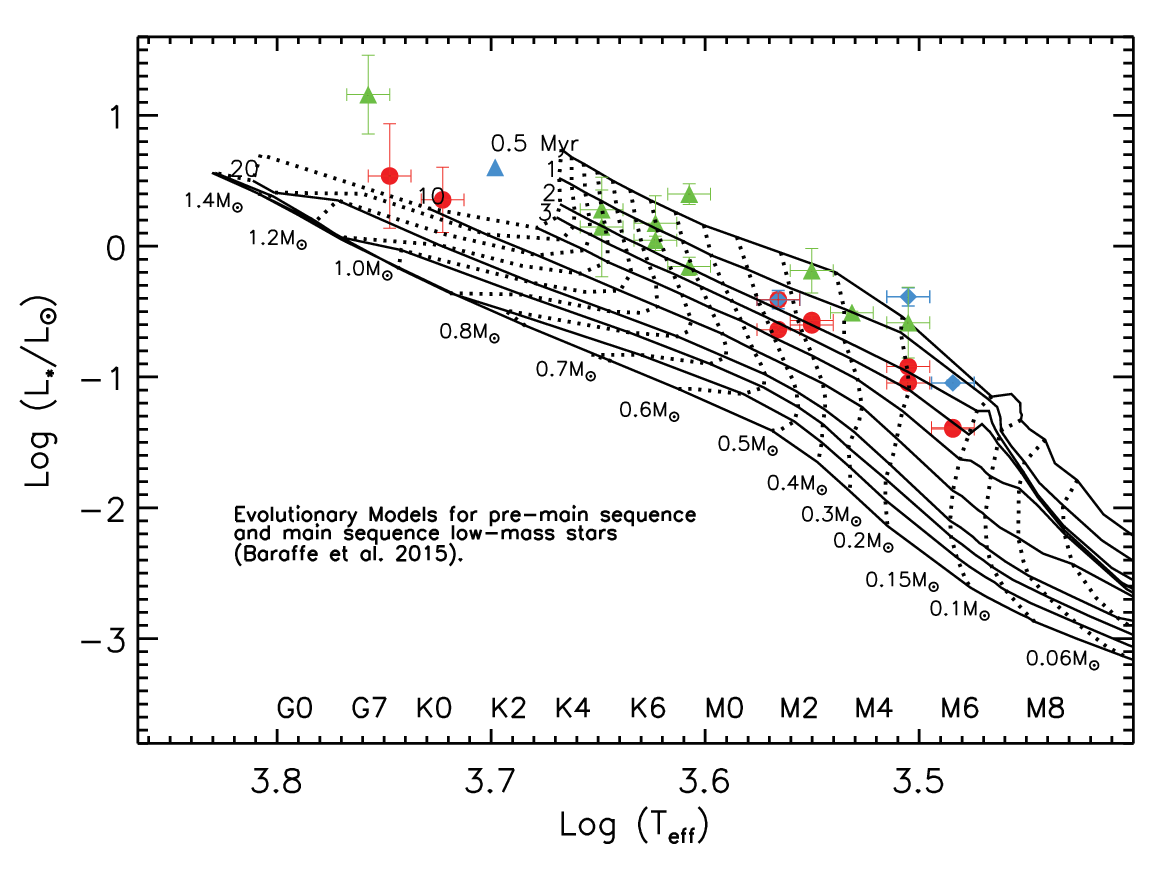} 
   \caption{Theoretical models from \citet{Baraffe2015} for low mass young stars. Solid lines in descending order are 0.5, 1, 2, 3, 5, 10, 20, 50 and 100 Myrs isochrones and dashed lines represent the evolutionary tracks in the range of 0.06 and 1.4 M. Blue solid diamonds, green solid triangles and red solid dots are TDs from Taurus-Aurigae, Ophiuchus and IC348 star forming regions, respectively. The blue solid triangle corresponds to EM* SR 24S with $L_{\odot}$ and T (K) taken from \citet{Andrews2011}. We used the scale temperatures range from \citet{Pecaut2013} and stellar luminosities are estimate as described in section ~\ref{sec:stellarparameter}}
 \label{fig:baraffe}
\end{figure*}

Distances to the star forming regions were adopted from literature. Thus, a distance of 140 $\pm$ 10 pc was adopted to the \textit{Taurus-Aurigae star forming region}  \citep{Loinard2008}. However, we found a large range of discrepancies in the distances to Ophiuchus and IC-348 members, leading us to carefully choose the most appropriate values, since these young members placed in the H$-$R diagram are sensitive to any variation of these distances (Figure ~\ref{fig:baraffe}). We adopted a distance to the \textit{IC 348 Region} based on the distance to the Perseus molecular cloud, that has been estimated in a wide range between 220 - 380 pc  \citep{Harris1954, Herbig1983, Cernis1993, Scholz1999, deZeeuw1999, Hirota2008,Hirota2010}. We examined in detail the distance to the brightest cluster member, LRL 1.  As a pair of B5V stars, these objects are zero-age main sequence (ZAMS) so they have a model-derived luminosity that is almost independent of age. Using the same technique as described above, we confirmed the distance of 220$\pm$10\,pc, excluding ZAMS model uncertainties but including uncertainties in reddening. We provide a more detailed discussion about the IC 348 distance in Appendix A. In the case of the distance to the \textit{Ophiuchus} region and as the position of most of our objects sit around the main cloud, L1688, we based our decision on the distance measured to this association. For our purposes and due to the wide extension in the position of the TDs, we adopted a mean distance of 130 pc to L1688 core \citep[and references therein]{Wilking2008} and consistent with the distance of 131 $\pm$ 3 pc calculated by \citet{Mamajek2008}.

Once the distances were found and constrained, we calculated the bolometric luminosities ($L_{\star}$) of IC 348, Ophiuchus and Taurus-Auriga members.  Estimations of stellar masses and ages are obtained from the H-R diagram and \citet{Baraffe2015} evolutionary tracks, except for IC348-21, IC348-31 and EM* SR 21A, where we made use of the PARSEC evolutionary models \citep{Bressan2012}. These values are estimated with the use of  L$_{\star}$ and K absolute magnitudes. Age and stellar mass uncertainties are based mainly on the H-R diagram placement and the determination of L$_{\star}$, since T$_{\rm eff}$ does not vary with a large magnitude at ages $<$ 5 Myrs in the pre-main sequence evolution tracks of low mass stars. The main sources of error on the  L$_{\star}$ uncertainties are distance and extinction \citep{Hartmann2001}. In our estimates of $A_{J}$, we used the expected colour of a main-sequence star, which underestimates these values up to a factor of 2, and it is reflected in the luminosity and age of the system. In the special case of embedded object EM* SR 24S that belongs to the triple system EM* SR24, we adopted stellar properties from \citet{Andrews2011} to avoid any IR contribution from its nearby binary companion, EM* SR 24N \citep[0".2;][]{Simon1995}. Table \ref{table:estimatedproperties} shows the estimated stellar parameters. Figure ~\ref{fig:baraffe} shows Baraffe evolutionary models with IC348, Taurus-Aurigae and Ophiuchus members. They are dispersed in a range of ages between 0.5 and 10 Myrs, characteristic of T-Tauri stars. Here, we carefully selected the chosen isochrone to derive stellar masses for each target, because these isochrones are also used in our Bayesian analysis, as explained in section ~\ref{sec:bayes}.

\begin{table*}
 \centering
 \begin{minipage}{114mm}
  \caption{Luminosity and Extinction for Our Sample Members.}
 \label{table:estimatedproperties}
  \begin{tabular}{lccccccccl}
\hline \hline \\[-2ex]
   \multicolumn{1}{c}{\textbf{Target}} &
   \multicolumn{1}{c}{\textbf{Temperature}\footnotemark[1]} &
   \multicolumn{1}{c}{\textbf{$A_{J}$\footnotemark[2]}} &
   \multicolumn{1}{c}{\textbf{Luminosity}}  &
   \multicolumn{1}{c}{\textbf{Age}}  &
   \multicolumn{1}{c}{\textbf{Mass}\footnotemark[3]} \\[0.8ex] 
   \multicolumn{1}{c}{$\mathbf{ }$} &
   \multicolumn{1}{c}{\textbf{[K]}} &
   \multicolumn{1}{c}{$\mathbf{[mag] }$} &
   \multicolumn{1}{c}{\textbf{$\mathbf{ [ \,\Lsun ]}$}} &
   \multicolumn{1}{c}{$\mathbf{[Myrs] }$} &
   \multicolumn{1}{c}{\textbf{$\mathbf{ [ \,\Msun ]}$}} &
   \multicolumn{1}{c}{\textbf{}} \\[0.5ex]\hline \hline \\[-5ex]    

\\[-0.8ex] 

\multicolumn{6}{c}{IC 348 d = 220 pc }\\
\\[-3.2ex]\hline

LRL 21 & 5280&1.70 $\pm$ 0.07&2.26 $\pm$ 0.25&5.0$_{-2.0}^{+3.0}$& 1.60$_{-0.05}^{+0.09}$\\
LRL 67 & 3680&0.82 $\pm$ 0.06&0.23 $\pm$ 0.03&3.8$_{-1.0}^{+1.2}$ & 0.50$_{-0.10}^{+0.05}$\\
LRL 72&3550&0.97 $\pm$ 0.08&0.27 $\pm$ 0.04&1.9$_{-0.5}^{+0.4}$& 0.41$_{-0.04}^{+0.07}$ \\
LRL 237&3050&0.47 $\pm$ 0.10 &0.04 $\pm$ 0.01&3.4$_{-0.6}^{+0.7}$&0.13 $_{-0.04}^{+0.02}$\\
LRL 97&3550&1.79 $\pm$ 0.08 &0.25 $\pm$ 0.03&2.0$_{-0.5}^{+0.8}$& 0.37$_{-0.04}^{+0.07}$\\
LRL 31&5590&3.21 $\pm$ 0.05 &3.44 $\pm$ 0.40 &8.0$_{-3.0}^{+2.0}$ & 1.62$_{-0.06}^{+0.09}$\\
LRL 182&3200&1.03 $\pm$ 0.07 &0.09 $\pm$ 0.01&2.7$_{-0.6}^{+0.4}$& 0.22$_{-0.03}^{+0.05}$\\
LRL 213&3050&0.53 $\pm$ 0.16 &0.04 $\pm$ 0.01&3.4$_{-0.6}^{+0.7}$& 0.13$_{-0.04}^{+0.02}$\\
LRL 58 &3680&1.13 $\pm$ 0.07 &0.39 $\pm$ 0.05&1.7$_{-3.0}^{+3.0}$ & 0.50$_{-0.07}^{+0.05}$\\
LRL 135& 3200&0.74 $\pm$ 0.11 &0.12 $\pm$ 0.02&1.9$_{-0.3}^{+0.4}$& 0.2$_{-0.03}^{+0.05}$\\[-0ex]\hline\\[-3ex]

\multicolumn{6}{c}{Taurus-Aurigae d = 140 pc }\\
\\[-3.2ex]\hline

IRAS04125+2902&3680&0.90 $\pm$ 0.06 &0.39 $\pm$ 0.07&1.7$_{-0.5}^{+0.4}$& 0.50$_{-0.02}^{+0.04}$\\
V410 X-ray 6&3200&0.98 $\pm$ 0.09 &0.41 $\pm$ 0.07&0.1$_{-0.1}^{+0.4}$ &0.22$_{-0.05}^{+0.04}$\\
 J04210934+2750368&3050&0.00&0.09 $\pm$ 0.02&1.4$_{-0.5}^{+0.4}$ & 0.17$_{-0.07}^{+0.02}$\\\\[-3ex]\hline\\[-3ex]

 \multicolumn{6}{c}{Ophiuchus d = 130 pc }\\
\\[-3.2ex]\hline

EM* SR 24S\footnotemark[4]& 4990 & 7.00 & 4.00 &--&  2.00 \\
EM* SR 21A & 5720  & 2.46 $\pm$ 0.12 & 14.40 $\pm$ 2.40 &2.0$_{-1.0}^{+2.0}$& 2.70 $_{-0.10}^{+0.10}$ \\ 
WSB 12&4450&0.96 $\pm$ 0.13 &1.40 $\pm$ 0.15 & 2.1$_{-0.9}^{+1.8}$& 1.11 $_{-0.11}^{+0.16}$\\
 J16262367-2443138&4450&1.14 $\pm$ 0.15 &1.90 $\pm$ 0.21&1.3$_{-0.7}^{+1.2}$& 0.99 $_{-0.10}^{+0.14}$\\
 J16273901-2358187&4200&1.49 $\pm$ 0.11 &1.51 $\pm$ 0.17& 0.9$_{-0.4}^{+0.9}$& 0.90 $_{-0.09}^{+0.11}$\\
WSB 63&3550&1.68 $\pm$ 0.08 &0.65 $\pm$ 0.07& 0.6$_{-0.2}^{+0.1}$& 0.35 $_{-0.10}^{+0.09}$\\
 J16335560-2442049&4050&1.26 $\pm$ 0.09 &0.70 $\pm$ 0.08& 2.0$_{-0.5}^{+0.8}$& 0.73 $_{-0.09}^{+0.11}$\\
 J16250692-2350502&3400&1.32 $\pm$ 0.09 &0.31 $\pm$ 0.03& 1.1$_{-0.2}^{+0.4}$& 0.28 $_{-0.14}^{+0.05}$ \\
 J16315473-2503238&4050&2.25 $\pm$ 0.18 &2.50 $\pm$ 0.27&0.1$_{-0.1}^{+0.7}$& 0.86 $_{-0.20}^{+0.20}$\\
WSB 40&4200&1.68 $\pm$ 0.08 &1.11 $\pm$ 0.12& 1.5$_{-0.5}^{+0.6}$& 0.96 $_{-0.08}^{+0.12}$\\
V*V852 Oph&3200&0.84 $\pm$ 0.09 &0.26 $\pm$ 0.03&  0.6$_{-0.1}^{+0.3}$& 0.19 $_{-0.03}^{+0.08}$ \\
\\[-3ex]

 \hline 
\end{tabular}
$^{1}$ References:  Effective Temperatures are taken from the scale of \citet{Pecaut2013}, with uncertainties of $\sim$ 150 K \\
$^{2}$ Extinctions are calculated following the \citet{Mathis1990} approach. \\
$^{3}$ Masses are estimated assuming single star systems.\\
$^{4}$ Stellar parameters taken from \citet{Andrews2011}.

\end{minipage}
\end{table*}

\subsubsection{Dust Temperature and Inner Radii}

 \begin{figure*}
 \centering
     \includegraphics[width=0.96\textwidth]{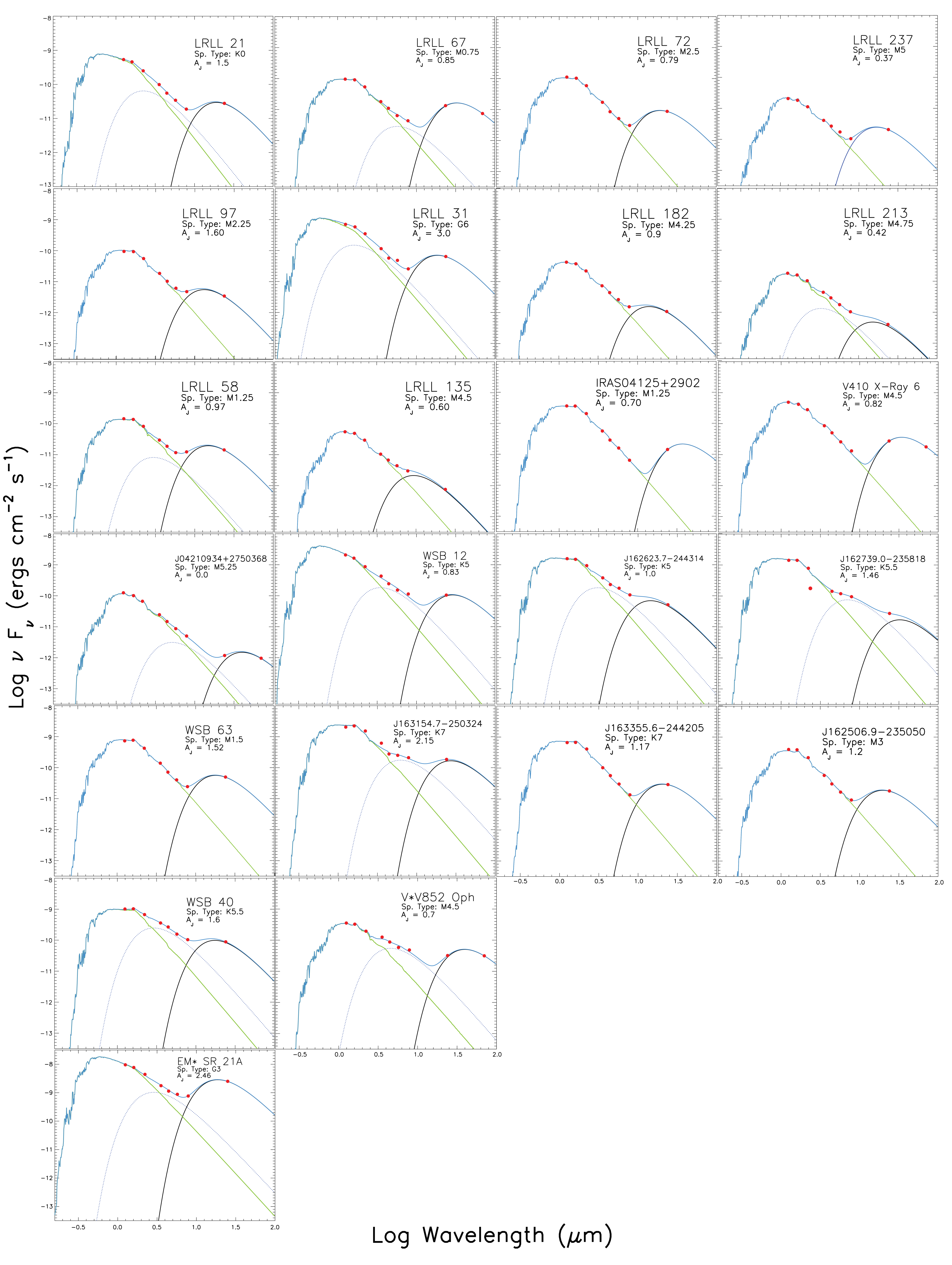}
             \caption{Spectral energy distribution of the sources classified as transitional disks. Red dots show photometric data acquired from the literature, green line is the BT-settl spectrum model according to the spectral type, blue line is the best fit, and the solid black and dotted blue lines are the disk black-body function values. $A_J$ values used are in Table ~\ref{table:estimatedproperties}. }
  \label{fig:imagessed} 
 \end{figure*}

As demonstrated by \citet{Espaillat2012}, the NIR excesses of classical TDs are well reproduced by the emission of a vertical wall directly exposed to stellar radiation. Then, we have computed the TD SEDs from NIR to MIR wavelengths in order to estimate inner wall radii for every disk (Figure \ref{fig:imagessed}). Since the thermal balance between emission and absorption of radiation is dominated by the dust grains as the main opacity source, we computed the dust temperature ($T_{d}$) at the truncation radius ($r_{d}$) of the disk.  Our input photometry for SED fitting were from 2MASS (1.25, 1.65, 2.22 $\mu$m ) and Spitzer/IRAC (3.6, 4.5, 5.8, 8 and 24 $\mu$m)  \citep{Skrutskie2006, Evans2003, Evans2009a, Currie2009, Rebull2010}.

An estimation of the $T_{d}$ was computed by fitting the stellar photosphere$+$disk black body function ($F_{m}$) to the observed data ($F_{\nu}$) and minimizing the sum of squares. Prior to this fitting, the photometric data were dereddened using the \citet{Mathis1990} approach, since the properties of the inner disk material significantly affects flux emission in these bands, and thus, the inner radii approximations \citep{Furlan2011}. To calculate the stellar synthetic photometry with a fixed temperature T$_{\star }$, which is approximated by the T$_{\rm eff}$, we 1) interpolated the response curves, for the set of filters used in the fitting, and the BT-Settl spectra models of the corresponding T$_{\star }$ \citep{Allard2014} and 2) convolved the filter response curves with the syntethic spectra, to match the spectral resolution. Because the 2MASS, IRAC, and 24 $\mu$m data have a photometric uncertainty of between a few percent and 0.1 mag for the objects investigated here, systematic effects can contribute up to 0.1 mag and also, to account for flux variability of the objects, we added an observational error of 12$\%$. A multiplicative dilution factor relating the central star radius ($R_{\star}$) and the distance to the object ($d$) is part of the minimization of the $\chi ^{2}$. Then, the model of received flux is the product of a dilution factor and the blackbody flux. In the case of a star, this dilution factor is given by $M_{d}=\left (\frac{R_{\star}}{d}  \right )^{2}$ \citep[e.g.][]{Bayo2008}

Finally, the  estimations of the truncation radius, $r_{d}$, are obtained by assuming Local Thermodynamic Equilibrium, and the ability of the dust to acquire thermal balance between absorption and emission of the radiation. The inner wall, nearly perpendicular to the stellar radiation, is heated only by the central star with a characteristic $R_{\star }$ and $T_{\star }$. Additionally, if the scattering of the dust grains is negligible and assuming optically thin gas in the inner region, we have in radiative equilibrium that the inner wall is being truncated at:

  \begin{equation}
 r_{d}=\frac{R_{\star }}{2\sqrt{}\varepsilon }\left ( \frac{T_{\star }}{T_{d}} \right )^{2}
 \end{equation}

 where $\varepsilon\equiv \frac{\kappa(T_{d}) }{\kappa(T_{\star }) }$ is the \textit{thermal cooling efficiency factor} that characterizes the dust properties of a certain size \citep{ Dullemond2010}. If the inner wall consist of small dust grains of radius $a$ $<<$ 3 $\mu$m and the backwarming by the grains deeper in the wall is negligible, then $\varepsilon$ $<<$ $3^{-0.5}$. For our purposes, the size of the grains at the location of the inner wall is taken to be $a$ $\sim$ 0.1  $\mu$m, leading to estimations of $r_{d}$ with $\varepsilon$ $\sim$ $0.08$ \citep[and references therein]{Isella2005} as shown in Table ~\ref{table:bayes}. For EM* SR 24S, because of its large inclination and elongated ring with a significant brightness asymmetry along the major axis, and to avoid any flux contamination in the Spitzer/IRAC bands from EM* SR 24N, we adopted the inner radius estimated by \citet{Andrews2011}, and recently confirmed by  \citet{VanderMarel2015}. Our $r_{d}$ estimations are in agreement with those previously measured (Table ~\ref{table:Properties}).

\section{Data Analysis}
\label{Sec:datanalysis}

\subsection{Bayesian Analysis}
\label{sec:bayes}

Although some secondary companions are observed in the interferograms, e.g. Figure ~\ref{fig:interferogram}, it is important to carefully account for the assumptions inherent in the imaging completeness. A conservative approach where only secure detections are considered and conservative detection limits are quoted does not make maximum use of the data, especially at the smallest separations where binaries of moderate contrast ratios give relatively small closure-phase signals. Here, we made use of Bayesian statistics to compute confidence levels for detections, providing the advantage of using prior information of the underlying population of faint stellar companions. In essence, our approach to completeness correction, along with extensive Monte Carlo simulations, assigns the probability of detecting the presence of a faint companion or absence of it. We built two hypotheses, binary ($B_{n}$) and single ($S_{n}$),  by using prior information of these models together with the aperture masking data of the TDs. Thus, a Bayes' theorem expresses the strengths of the hypotheses as follows:

\begin{equation}
p(B|D)=\frac{p(B_{n})p(D|B_{n})}{p(D)}
\label{eq:bayes2}
\end{equation}

\begin{equation}
p(S|D)=\frac{p(S_{n})p(D|S_{n})}{p(D)}
\label{eq:bayes1}
\end{equation}

where $p(D|B_{n})$ and $p(D|S_{n})$ are the \textit{global likelihood}\footnote{The global likelihood of a model is equal to the weighted average likelihood for all the parameters in consideration.} functions or probability of obtaining data D, if $B_{n}$ or $S_{n}$ are true; $p(B_{n})$ and $p(S_{n})$ are the prior probabilities; and $p(B_{n}|D)$ and $p(S_{n}|D)$ are the posterior probabilities of $B_{n}$ and $S_{n}$, respectively. The index $n$ represents the number of simulations performed to determine the probability for each model.

\subsubsection{Global Likelihood: Confirming or Ruling out the Presence of a Binary System.}

In the data reduction process, after computing closure phases, the calibrated data set is used to search for faint companions close to the central star and orbiting the inner region of the disk. Our search strategy is based on the computation of the global likelihood that can be maximized from the \textit{conditional likelihood} and its \textit{joint prior probability} (Equation \ref{eq:global}). Here, the conditional likelihood expresses the probability of observing our data for a specific set of model parameters and is weighted by the joint prior probability that incorporates prior information about the distribution of the model parameters. In our case, we have two hypotheses, $B_{n}$ and $S_{n}$, that can be tested by computing their global likelihood as follows:

\paragraph{Confirming a Binary System:} The maximum global likelihood or the odds by which our data favors a binary model, lies in our approach to completeness correction and the extensive binary star simulations to assign all possible contrasts of the secondary relative to the primary ($C$), angular separation ($\rho$) and position angle ($\theta$) values, as detailed below.  Thus, our analysis to compute confidence levels is based on the derivation of the $\chi^{2}$ goodness-of-fit to the $n$ mock binary system models. From Equation \ref{eq:bayes2} and marginalizing over all possible model parameters, we have that the global likelihood for a binary system model is:

\begin{equation}
P(D|B_{n})=\int d\Psi  \; p(\Psi |B_{n})\times P(D|B_{n},\Psi )
\label{eq:global}
\end{equation}

where, $ P(D|B_{n},\Psi )$ is the \textit{conditional likelihood}, $P(\Psi |B_{n})$ is the \textit{joint prior probability} for the model parameters and $\Psi = (T, a, e,\Omega,\omega, i,t, q)$ is the eight-dimensional parameter space representing all the possible binary orbits with the orbital parameters: Time of periastron passage ($T$, years), Semi-major axis ($a$, arc-second), Eccentricity ($e$),  Position angle of the line of nodes ($\Omega$), Longitude of periatron ($\omega $), Inclination ($i$, degrees), Epoch of observation ($t$, year) and Mass ratio ($q$). 

To compute the eight-dimensional integral shown in Equation \ref{eq:global}, we made use of \textit{Monte Carlo Integration}  by generating a number of random samples according to a determined probability distribution function (PDF) in a specific spatial domain. The prior PDFs for $T,\Omega$ and $\omega$ for a binary system are assumed to have a uniform distribution (uninformative prior), so $ p   ( T,\Omega,\omega_{2}|B_{i})=1$ in the space domain of these parameters. The inclination $i$ is sampled considering that the orbital plane of the generated binaries might have any orientation in space.  For $q$, we considered the mass ratio distribution often modelled in the form $f(q)\propto q^{\beta}$, where $\beta$ is a power index of 0 that reasonably describes our TD sample with spectral types ranging G3-M5  \citep[e.g.][]{Janson2012, Raghavan2010}. For the prior distribution of $e$ for these very young objets in a process of orbital circularisation, we based our approach on our resolving limitations of up to $\gtrsim$ 2 au that corresponds to orbital periods of  $\sim$ 1000 days and is sampled well enough by the ``thermal" eccentricity distribution $f(e) = 2e$ \citep{Ambartsumian1937, Duquennoy1991}. For simplicity, we used the logarithmically flat distribution $dN/d\,$log$\,a\propto a^{0}$ approach used by \citet{Metchev2009} to sample $a$. Thus, the joint prior probability is:

\begin{equation}
P (\Psi|B_{n})\propto\; q^{\beta}\;  \left ( \frac{2e}{a} \right )\;  \sin(i).
\end{equation}

Table ~\ref{table:range} shows the limits used in the simulations and corresponding prior for the orbital parameters. The sampling ranges for $i$, $e$, $T,\Omega$ and $\omega$ were taken by considering the total orbital plane with any orientation in space. More relevant for our detection limits, we focus our search region, mostly constrained by $a$, because the dynamics of binary-disk interaction models predict that tidal interaction between the binary star and the disk might truncate the inner region of the disk at radii of 2-3 times the semi-major axis of the binary orbit \citep{Artymowicz1994}. Table \ref{table:bayes} shows the inner radii estimates that we used as limits to sample the semi-major axis space that ranges between $\frac{r_{d}}{2}$ and $\frac{r_{d}}{3}$ at distances of the corresponding star-forming region (Tau-Aur: 140 pc, Ophiuchus: 130 pc and IC348: 220 pc).

Separately, Contrast ratio ($C$) and Stellar Masses ($M_{1}, M_{2}$) are sampled as a function of the corresponding distribution of the  Mass ratio ($q$), where these stellar masses together with the sample of semi-major values are used as inputs in Kepler's third law to compute the Orbital Periods (P, year) of each system. Here, we provided a brief summary of our methodology to sample $M_{1}, M_{2}$ and $C$.

\begin{itemize}

\item{\textit{Stellar Masses and Contrast Ratios:}} Initially, we chose an isochrone ($Z_{1}$) from the \citet{Baraffe2015} models, which represents the age of the TD and that in principle, plays the role of the primary component. In order to account for all the possible flux ratios, the following step is to start computing evolutionary tracks that describe the possible secondary stellar component in our data. This is done by randomly sampling a mass-ratio ($q_{n}$) distribution and using the relation $Z_{2,n}=  Z_{1} q_{n}$, where $Z_{2,n}$ and $Z_{1}$ are the tracks for all the possible secondary stars and the track for the primary star, respectively. Then, we computed the theoretical magnitudes of the these new evolutionary tracks by interpolating onto the theoretical stellar masses and K magnitudes of the chosen isochrone ($Z_{1}$). Thus, we have generated a series of isochrones that correspond to every value of $q_{n}$, $q=0$ being a single star and  $q=1$ a binary star system with similar masses. Once the evolutionary isochrones describing all the secondary components are computed and taking $Z_{1}$, we were able to calculate the total K magnitude of the binary system. Then, we estimated the primary mass ($M_{1}$) based on its observed absolute K magnitude interpolated onto the total K magnitude of the binary system and the theoretical mass track; and with the relationship $M_{2}= q_{1,n}M_{1}$, we obtained the secondary mass, $M_{2}$. Finally, interpolating the primary and secondary masses onto the total K magnitude of the binary system and the theoretical mass track, we estimated their K magnitudes in order to compute contrast ratios. 

\end{itemize}

Finally, the orbital parameters sampled are used to derive the angular separation ($\rho$) and position angle ($\theta$) \citep{Meeus1992} for N simulated binary systems. These values together with the calibrated data are used to compute the maximum likelihood of the contrast ratio and, thus confirm or rule out a stellar companion orbiting in the inner region of the disk.

\begin{table}
 \centering
 \scriptsize

 \begin{minipage}{100mm}
  \caption{Parameter Prior for Binary Model}
 \label{table:range}
  \begin{tabular}{lccc}
\hline \hline \\[-2ex]
   \multicolumn{1}{c}{\textbf{Orbital Parameter}} &
   \multicolumn{1}{c}{\textbf{Prior}\footnotemark[1]} &
   \multicolumn{1}{c}{\textbf{Lower Bound}} &
   \multicolumn{1}{c}{\textbf{Upper Bound}} \\[-1ex] 

\multicolumn{4}{c}{}\\
\\[-3.2ex]\hline

Semi-major Axis (arc second)&Jeffreys' prior&$\frac{1}{3}R_{d}$&$\frac{2}{3}R_{d}$\\
Period (years)   & Keplearian&$a_{i}^{3}$&$a_{o}^{3}$\\
Time of periastron passage (years)  & Uniform & --&-- \\  
Eccentricity&2$e$ &0.01&0.9 \\
Inclination (degree)&$\sin$ (i) &0&180\\
Node (degree)&Uniform&0&180\\
Longitude of Periastron (degree)&Uniform&0&360\\
Mass-ratio &Power-Law&0.01&1\\
 \hline 
 \\[-3ex]
\end{tabular}
\\
$^{1}$  It was used the Cumulative Distribution Function to sample in the orbital parameter space. 
\end{minipage}
\end{table}

\paragraph{Confirming a Single System:} In the case of a single system as a point-symmetric target, the calibrated closure-phases are nearly equal to zero. Then, the computation of the global likelihood for a single system is basically determined by the conditional likelihood for a single system model with flux ratio equal to zero. From Equation \ref{eq:bayes1}, the global likelihood is:

\begin{equation}
p   ( D|S_{n}) = p   ( C=0|S_{n}).
\end{equation}

\paragraph{Odds Ratio:} 

After computing the global likelihoods for $B_{n}$ and $S_{n}$, we are interested in comparing the two hypotheses, thus we computed the ratio of $p(B|D)$ and $p(S|D)$ known as \textit{Odds Ratio} and written as $O_{B,S}$ in favor of the binary system model over the single system model:

\begin{equation}
O_{B,S}=\frac{p(B|D)}{p(S|D)}=\frac{p(B_{n})}{p(S_{n})}\frac{p(D|B_{n})}{p(D|S_{n})}
\label{eq:odds}
\end{equation}

where the first factor on the right side of Equation \ref{eq:odds} corresponds to the \textit{prior odds ratio} equal to unity due to its uniformity in the parameter space and the second term is known as \textit{Bayes' Factor} ($\Upsilon$). Thus, $\Upsilon_{B,S}>> $ 1 means that a binary model is preferred by the data, $B_{B,S}<< $ 1 the data comes from a single star and $B_{B,S}\approx $ 1 means that the odds are not modified and a binary and/or single star are equally probable.

\section{Results}

\begin{table*}
 \centering
 \begin{minipage}{160mm}
  \caption{Companions Identified Outside the Inner Radii with the Aperture Mask}
 \label{table:nosample}
  \begin{tabular}{lcccccccc}
\hline \hline \\[-2ex]
   \multicolumn{1}{c}{\textbf{Primary}} &
   \multicolumn{1}{c}{\textbf{BJD}} &
   \multicolumn{1}{c}{\textbf{$\Delta$K}} &
   \multicolumn{1}{c}{\textbf{Sep}} &
   \multicolumn{1}{c}{\textbf{Sep}} &
   \multicolumn{1}{c}{\textbf{PA}} &
   \multicolumn{1}{c}{\textbf{M$_{1}$\footnotemark[1]}} &
    \multicolumn{1}{c}{\textbf{M$_{2}$\footnotemark[1]}} &
    \multicolumn{1}{c}{\textbf{Sig.}\footnotemark[2]} \\[0.2ex] 
   \multicolumn{1}{c}{$\mathbf{ }$} &
   \multicolumn{1}{c}{$\mathbf{ (2400000 +)}$} &
   \multicolumn{1}{c}{\textbf{[mag]}} &
   \multicolumn{1}{c}{$\mathbf{ [mas]}$} &
   \multicolumn{1}{c}{$\mathbf{[au] }$} &
   \multicolumn{1}{c}{$\mathbf{[deg.] }$} &
   \multicolumn{1}{c}{\textbf{$\mathbf{ [ \,\Msun ]}$}} &
   \multicolumn{1}{c}{\textbf{$\mathbf{ [ \,\Msun ]}$}} \\[-0.6ex]\hline \hline 
\multicolumn{8}{c}{}\\
\\[-6.2ex]

LRL 72&55880.79&1.37 $\pm$ 0.02 & 103.72 $\pm$ 0.36 & 22.82 $\pm$ 0.04 & 221.86 $\pm$ 0.10&0.41&0.13& 49.0\\
"&56883.04  &1.42 $\pm$ 0.03 &  100.45 $\pm$ 0.47  &  22.10 $\pm$ 0.08 & 226.70 $\pm$ 0.26&''&''&40.0\\
LRL 182&57000.75&1.54 $\pm$ 0.04 &  35.30 $\pm $0.74 &  7.76 $\pm$ 0.26 & 213.42 $\pm$ 0.77&0.20&0.05&39.4\\
LRL 135&57000.79&0.17 $\pm$ 0.01 &  82.01 $\pm$ 0.26 &  18.04 $\pm$ 0.01 & 208.36 $\pm$ 0.21&0.18&0.16&90.5\\
LRL 213 &57000.83&0.57 $\pm$ 0.18 &  18.01 $\pm$ 0.76 &  3.96 $\pm$ 0.17 &  50.48 $\pm$ 3.23&0.11&0.06&9.30\\
J04303235+3536133\footnotemark[3]  &55880.96& 0.40 $\pm$ 0.36 &  20.25 $\pm$ 2.88 & 9.11$\pm$ 1.30 & 205.28 $\pm$ 2.25& 0.72& 0.59&10.31\\
"     & 57001.00& 0.02 $\pm$ 0.02 &  26.24 $\pm$ 0.70& 11.81 $\pm$ 0.32 & 216.81 $\pm$ 2.19 &''& ''& 9.48\\
 \hline 
\end{tabular}
\\
$^{1}$ The fractional uncertainties on the individual masses are $\leq$20$\%$.\\
$^{2}$ Significance in $\sigma$ is calculated as $\sqrt{\Delta\chi^2 \times (N_{\rm df}/N_{\rm cp})}$, with $N_{\rm df}$ the number of degrees of freedom and  $N_{\rm cp}$ the number of closure-phases, and uncertainties scaled so that the reduced chi-squared of the best fit solution is unity. \\
$^{3}$ Located at Auriga-California molecular cloud \citep[450 $\pm$ 23 pc;][]{Lada2009}.
\end{minipage}
\end{table*}

We are considering visual companions to stars in our cluster down to a Ks magnitude of $\sim$15 in IC348 (and significantly brighter limits in other clusters), where the stellar density is approximately $2 \times 10^{-4}$ stars per square arcsec at this magnitude limit. Given that we are only considering companions within $\sim$0.16 arcsec in this paper, the probability of a background star masquerading as a physical companion is no more than 0.002\% for each star in the sample, or $<$0.04\% for the entire sample.

As a part of our analysis and election of calibrator objects, we have found that 2MASS J04303235+3536133 is a binary object and although, it is reported as a TD by \citet{Cieza2010}, we excluded this object from the sample because it is located at the Auriga-California molecular cloud. Table~\ref{table:nosample} shows contrast and positions for both observing epochs of the object.

\subsection{Stellar Companions Outside the Inner Radii}
\label{Sec:companionoutside}

In our first identification process, we identified LRL 72, LRL 182, LRL 213 and LRL 135 as cases for a high contrast detection with a confidence level of $\geq$ 99.5 $\%$ (section \ref{Sec:Target}). We removed those objects as a part of the sample of TDs because the stellar companion is located outside the area of study and is not responsible for carving out the inner region of the disk. Estimation of the companion mass values were obtained by taking the K mag from the system,  $\Delta$K and \citet{Baraffe2015} models. Then, after obtaining an estimate of the K magnitudes for each stellar component, we interpolated linearly onto the evolutionary tracks to obtain the individual masses. Mass uncertainties are dominated by the evolutionary tracks, and depending on the theoretical model used, are of the order of 10-20$\%$ \citep{Siess2001}. In the case of LRL 213, we included visibility amplitudes for breaking contrast/separation degeneracies. Table~\ref{table:nosample} shows the epoch, contrast ratio detection, position parameters and companion masses of these new binary systems.

\begin{figure}[!h]
 \centering 
    \includegraphics[width=0.48\textwidth]{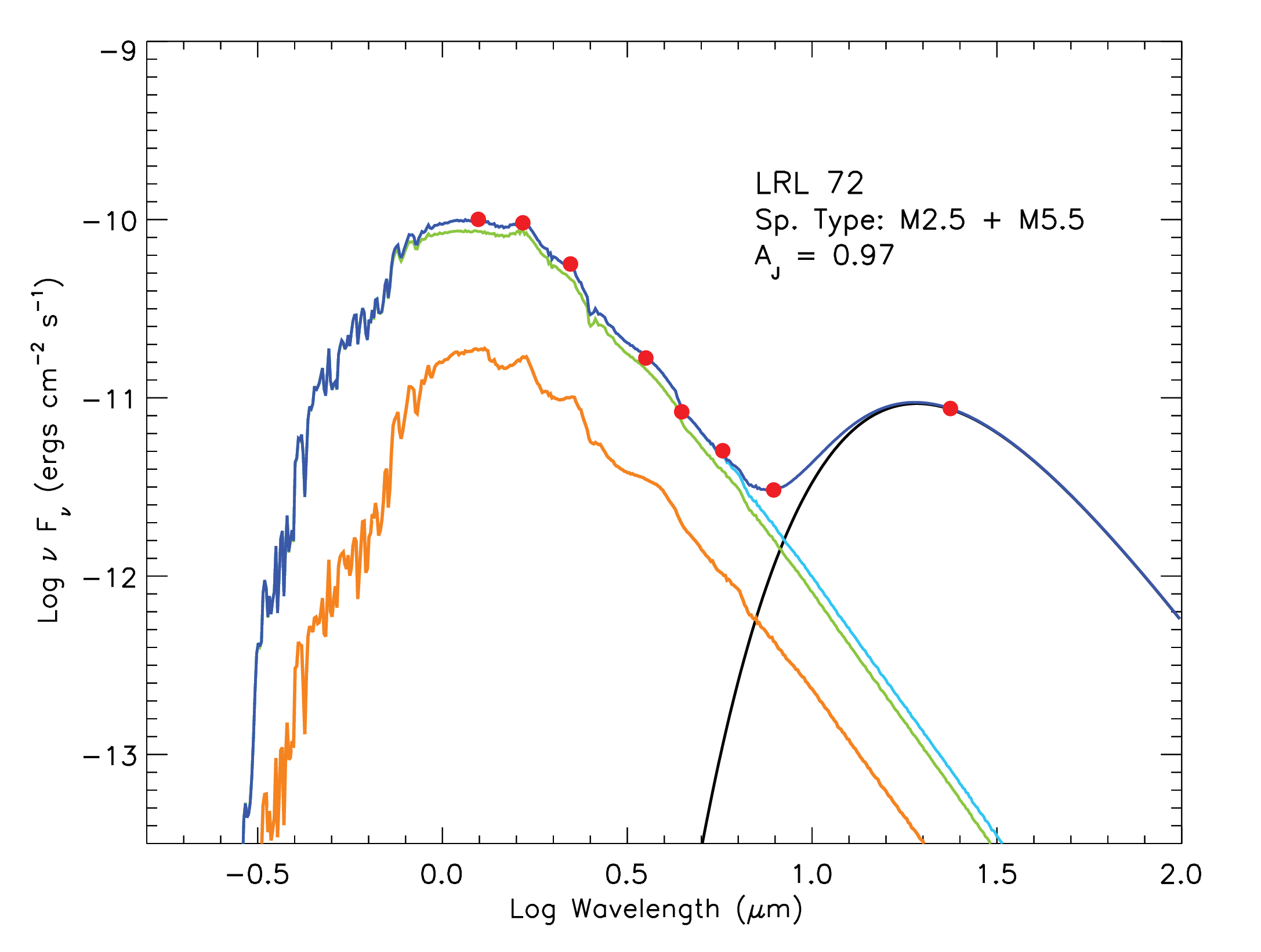}
 \caption{ Spectral Energy Distribution for LRL 72 and its components with spectral types M2.5 and M5.5. Red dots show photometric data acquired from the literature, green and orange lines are the primary and secondary components BT-settl spectra, respectively. The black line is the disk black-body function values, the cyan line is the best stellar fit and the blue line is the sum of the primary and secondary spectra and the disk black-body function values. }
 \label{fig:twospectra}
\end{figure}

The SEDs of these four close binary companions have NIR and MIR fluxes above the level expected for a reddened low mass star of spectral types M2-M5. The qualitative SEDs of these objects can be explained by a system composed of a close low-mass binary star, where the secondary component is surrounded by a ``cold and weak" disk, a \textit{``circum-secondary disk"}. For instance, if the LRL 72 components are coeval and using the \citet{Baraffe2015} models, the object would correspond to a spectral type of $\sim$ M5$-$M7 with a temperature between 2900 and 3100 K. As shown in Figure \ref{fig:twospectra}, the IR excess in the SED might be emitted by a disk orbiting the secondary component, instead of the primary star. Previous studies have shown that disks around \textit{brown dwarfs and very low mass stars} are generally flatter and less massive than their counterpart T Tauri disks \citep{Olofsson2013, Liu2015}. This would explain the relative weak IR excesses observed in their SEDs of these binary systems.

\restylefloat{figure}

Additionally, the angular separation of $\sim$ 100 $\pm$ 0.4 mas  of LRL 72, imposes a limit on the extension of the disk of around 22 au, because the presence of the companion at this close distance would truncate the disk. Similarly, LRL 182 and LRL 213 with a closer companions having mass values of 0.07 $M_{\odot}$ at $\sim$ 7.7 au and 0.15 $M_{\odot}$ at  $\sim$ 4.0 au, respectively, the disk also undergoes a faster dispersion as shown in its SED with a homogeneously small IR excess, see Figure \ref{fig:imagessed}. For LRL135, the mass ratio is near unity ($q$=0.89), and the system has a very weak excess with a disk  (Figure ~\ref{fig:imagessed}). This system therefore remains a TD candidate, with it being unclear which component of the binary is the TD candidate. This system has a relatively high disk to stellar temperature ratio, with only a very small cleared inner disk region.

\subsection{Stellar Companions Inside the Inner Radii}
\label{Sec:Inner}

\begin{table*}
 \centering
 \begin{minipage}{102mm}
  \caption{Bayesian Analysis}
 \label{table:bayes}
  \begin{tabular}{lcccc}
\hline \hline \\[-2ex]
   \multicolumn{1}{c}{\textbf{Object}} &
   \multicolumn{1}{c}{\textbf{Bayes' Factor}} &
   \multicolumn{1}{c}{\textbf{$r_{d}$}} &
   \multicolumn{1}{c}{\textbf{$r_{d}$}} &
   \multicolumn{1}{c}{\textbf{T$_{d}$}} \\[-0.2ex] 
   \multicolumn{1}{c}{$\mathbf{ }$} &
   \multicolumn{1}{c}{$\mathbf{ }$} &
   \multicolumn{1}{c}{\textbf{[au]}} &
   \multicolumn{1}{c}{$\mathbf{ [mas]}$} &
   \multicolumn{1}{c}{$\mathbf{[K] }$}  \\[-0.6ex]\hline \hline 
\multicolumn{5}{c}{}\\
\\[-6.2ex]

LRL 21  & 0.03&10.9 $\pm$  0.8&49.0 $\pm$ 4.0&194.2 $\pm$ 1.0\\
LRL 67  & 0.03& 11.6 $\pm$ 1.1&53.0 $\pm$ 5.0 &112.0 $\pm$ 2.3\\ 
LRL 72\footnotemark[1]&-- &3.8 $\pm$ 0.4&17.0 $\pm$ 2.0&191.3$\pm$ 1.7\\
LRL 237&1.0&1.2 $\pm$ 0.1 &5.0  $\pm$ 0.4&219.6 $\pm$ 5.0\\
LRL 97&0.98 &1.7 $\pm$ 0.2&8.0 $\pm$ 1.0&279.1 $\pm$ 3.4\\
LRL 31&300&13.1 $\pm$ 1.1 &60.0 $\pm$ 5.0&198.0 $\pm$ 4.6\\
LRL 182\footnotemark[1]&--&1.2 $\pm$ 0.1&6.0 $\pm$ 0.4&260.1$\pm$ 2.7\\
LRL 213&--&0.9 $\pm$ 0.1&4.0 $\pm$ 0.4&242.4 $\pm$ 1.0 \\
LRL 58 &0.99&2.8 $\pm$ 0.3& 13.0 $\pm$ 1.0&246.5 $\pm$ 1.4\\
LRL 135&0.99&0.6 $\pm$ 0.1&2.0 $\pm$ 0.4&393.0 $\pm$ 2.4\\
IRAS04125+2902&0.02&19.9 $\pm$ 2.0&143.0 $\pm$ 14.0 &92.3 $\pm$ 1.3 \\
V410 X-ray 6&$>$ 300&5.4 $\pm$ 0.4&39.0 $\pm$ 2.0 &183.4 $\pm$ 9.6\\
 J04210934+2750368&24&10.5 $\pm$ 1.2&75.0  $\pm$ 9.0 &91.0 $\pm$ 1.9\\
EM* SR 24S\footnotemark[2] & 0.01 & 29.0  & 223.0& -- \\
EM* SR 21A  & 0.003 & 27.2  $\pm$ 2.7 & 209.0 $\pm$ 21.0 &196.02 $\pm$ 9.8 \\
WSB 12&$>$ 300&6.0 $\pm$ 0.5& 46.0 $\pm$ 4.0&230.0 $\pm$ 10.7\\
 J16262367-2443138&0.07&8.3 $\pm$ 0.7&64.0 $\pm$ 5.0 &211.0 $\pm$ 2.3\\
J16273901-2358187&0.05&11.1 $\pm$ 1.0&85.0 $\pm$ 7.0 &174.2 $\pm$ 1.9\\
WSB 63&0.05&5.3 $\pm$ 0.5& 41.0 $\pm$ 4.0 &204.3 $\pm$ 2.0\\
 J16335560-2442049&$>$ 300&7.3 $\pm$ 0.7&56.0 $\pm$ 5.0 &178.2 $\pm$ 3.3\\
 J16250692-2350502 &1.0&4.4 $\pm$ 0.5&35.0 $\pm$ 4.0 &186.4 $\pm$ 3.3\\
 J16315473-2503238\footnotemark[3] &$>$ 300&6.2 $\pm$ 0.6&47.0 $\pm$ 4.0 &266.4 $\pm$ 11.5\\
WSB 40\footnotemark[4]&$>$ 300&6.7 $\pm$ 0.6&52.0 $\pm$ 4.0 &209.8 $\pm$ 1.4\\
V*V852 Oph&0.05&16.2 $\pm$ 2.1&125.0 $\pm$ 16.0 &93.1 $\pm$ 3.6\\
 \hline 
 
\end{tabular}
\\
$^{1}$ Targets not included in the statistical analysis of TDs.\\
$^{2}$ Estimated inner radius taken from \citet{Andrews2011}.\\
$^{3}$ Based on \citet{Kohn2016}. The Bayes' factor based on closure-phase alone was 30. See section \ref{Sec:Inner} for a detailed discussion.\\
$^{4}$ Using visibility amplitude. The Bayes' factor based on closure-phase alone was 3. See section \ref{Sec:Inner} for a detailed discussion.

\end{minipage}
\end{table*}

\begin{table*}
 \centering
 \begin{minipage}{100mm}
  \caption{Detection Confidence Limits (99.9$\%$)}
 \label{table:nondetections}
  \begin{tabular}{@{}lcccccc@{}}
\hline \hline \\[-2ex]
   \multicolumn{1}{c}{\textbf{Object}} &
   \multicolumn{1}{c}{\textbf{BJD}} &
   \multicolumn{1}{c}{\textbf{}} &
   \multicolumn{1}{c}{\textbf{}} &
   \multicolumn{1}{c}{\textbf{$\Delta$K\footnotemark[1]}} &
   \multicolumn{1}{c}{\textbf{}} &
    \multicolumn{1}{c}{\textbf{}} \\[0.2ex] 
   \multicolumn{1}{c}{\textbf{}} &
   \multicolumn{1}{c}{\textbf{(2400000 +)}} &
   \multicolumn{1}{c}{\textbf{10-20}} &
   \multicolumn{1}{c}{\textbf{20-40}} &
   \multicolumn{1}{c}{\textbf{40-80}} &
   \multicolumn{1}{c}{\textbf{80-160}} &
   \multicolumn{1}{c}{\textbf{160-240}} \\[-0.6ex]\hline \hline 

\\[-3.2ex]
\multicolumn{7}{c}{Non-Detections}\\
\\[-3.2ex]\hline
LRL 21  & 55880.75 &  1.13 &  3.13 &  3.99 &  3.72 &  3.77 \\
LRL 67  & 55880.83 &  1.83 &  3.59 &  4.44 &  4.18 &  4.24 \\
LRL 237&56882.04 & - &  0.23 &  1.52 &  1.00 &  1.14 \\
LRL 97&56882.04 &  0.59 &  2.75 &  3.80 &  3.54 &  3.57 \\
LRL 58 &57000.79 &  2.09 &  3.77 &  4.72 &  4.44 &  4.48 \\
IRAS04125+2902&56881.12 &  1.60 &  3.41 &  4.29 &  4.04 &  4.09 \\
EM* SR 24S & 54635.75 & - &  1.43 &  2.77 &  2.40 &  2.41  \\
EM* SR 21A  & 54635.88 &  2.51 &  4.11 &  4.98 &  4.71 &  4.75  \\
 J16262367-2443138& 54635.88 &  2.28 &  3.91 &  4.81 &  4.57 &  4.58 \\
J16273901-2358187&55674.96 & - &  0.44 &  2.01 &  1.60 &  1.67 \\
WSB 63& 55674.96 &  0.88 &  3.00 &  3.97 &  3.77 &  3.82\\
  J16250692-2350502 &55674.96 & - &  0.37 &  1.81 &  1.42 &  1.51 \\
V*V852 Oph&56032.08 &  1.60 &  3.41 &  4.31 &  4.06 &  4.12 \\
\\[-3.2ex]\hline
\multicolumn{7}{c}{Detections Outside Inner Radii}\\
\\[-3.2ex]\hline
LRL 72  & 55880.79 & - & - &  0.55 &  0.31 &  0.38 \\
''& 56883.04 & - & - &  0.54 &  0.31 &  0.38\\
LRL 182& 57000.75 & - &  0.13 &  1.08 &  0.70 &  0.80\\
LRL 135& 57000.79 & - & - &  0.20 & - &  0.03\\
LRL  213& 57000.83 &  0.68 &  2.85 &  3.83 &  3.63 & -\\
\\[-3.2ex]\hline
\multicolumn{7}{c}{Detections Inside Inner Radii}\\
\\[-3.2ex]\hline
LRL 31  & 57000.79 &  0.93 &  3.02 &  3.91 &  3.72 & -0.00 \\
V410 X-ray 6  & 55139.00 &  0.57 &  2.72 &  3.63 &  3.38 &  3.44 \\
WSB 12 & 55675.00 &  0.56 &  2.69 &  3.63 &  3.37 &  3.41\\
WSB 40 & 56032.08 &  1.45 &  3.32 &  4.33 &  4.06 &  4.11\\
J16335560-2442049 &  55675.04 &  0.93 &  3.02&  3.91 &  3.72& -\\

\hline 
\end{tabular}
\\
 NOTES: Angular separation ranges are given in mas.\\
$^{1}$  Limits within annuli.\\
\end{minipage}
\end{table*}

\begin{table*}
 \centering
 \begin{minipage}{140mm}
  \caption{Companions Identified Inside the Inner Radii with the Aperture Mask}
 \label{Table:parameters}
  \begin{tabular}{lccccccc}
\hline \hline \\[-3ex]
   \multicolumn{1}{c}{\textbf{Primary}} &
   \multicolumn{1}{c}{\textbf{BJD}} &
   \multicolumn{1}{c}{\textbf{$\Delta$K}} &
   \multicolumn{1}{c}{\textbf{Sep}} &
   \multicolumn{1}{c}{\textbf{Sep}} &
   \multicolumn{1}{c}{\textbf{PA}} &
      \multicolumn{1}{c}{\textbf{M$_{2}$\footnotemark[1]}} &
    \multicolumn{1}{c}{\textbf{Sig.}} \\[0.2ex] 
   \multicolumn{1}{c}{$\mathbf{ }$} &
   \multicolumn{1}{c}{$\mathbf{ (2400000 +)}$} &
   \multicolumn{1}{c}{\textbf{[mag]}} &
   \multicolumn{1}{c}{$\mathbf{ [mas]}$} &
   \multicolumn{1}{c}{$\mathbf{[au] }$} &
   \multicolumn{1}{c}{$\mathbf{[deg.] }$} &
   \multicolumn{1}{c}{\textbf{$\mathbf{ [ \,\Msun ]}$}} \\[-0.4ex]\hline \hline 
\\[-3.2ex]

LRL 31&57000.92&3.92 $\pm$ 0.20 &  38.09 $\pm$ 5.30 & 8.38 $\pm$ 1.16 & 45.56 $\pm$ 4.06&0.20&6.70\\
V410 X-ray 6&55138.92&0.15 $\pm$ 0.07 &  22.96 $\pm$ 1.25 &  3.22 $\pm$ 0.18 &87.80 $\pm$ 2.20&0.21&12.50\\
WSB 12&55674.92&0.42 $\pm$ 0.11 &  20.29 $\pm$ 0.78  & 2.64 $\pm$ 0.10 & 351.02 $\pm$ 2.13&0.75&11.80\\
WSB 40&56031.51&0.35 $\pm$ 0.14 &  17.42 $\pm$ 0.94  & 2.26 $\pm$ 1.12 & -11.37 $\pm$ 3.80&0.75&8.40\\
J16335560-2442049&55675.04&0.10 $\pm$ 0.05 &  25.30 $\pm$ 0.55& 3.29 $\pm$ 0.71 &  344.55 $\pm$ 1.77& 0.61 & 14.40\\
 \hline 
 \end{tabular}
 \\
NOTES: Significance in $\sigma$ is calculated as $\sqrt{\Delta\chi^2 \times (N_{\rm df}/N_{\rm cp})}$, with $N_{\rm df}$ the number of degrees of freedom and  $N_{\rm cp}$ the number of closure-phases, and uncertainties scaled so that the reduced chi-squared of the best fit solution is unity. \\
$^{1}$The fractional uncertainties on the stellar masses are $\leq$20$\%$.\\
\end{minipage}
\end{table*}

\begin{table*}
 \centering
  \begin{minipage}{130mm}
  \caption{Degenerate Companion Solutions for 2MASS J04210934+2750368 at $\Delta$K  0.5, 1, 2 and 3.}
 \label{table:reported}
  \begin{tabular}{ccccccccc}
\hline \hline \\[-3ex]
   \multicolumn{1}{c}{\textbf{Primary}} &
   \multicolumn{1}{c}{\textbf{BJD}} &
   \multicolumn{1}{c}{\textbf{$\Delta$K}} &
   \multicolumn{1}{c}{\textbf{Sep}} &
   \multicolumn{1}{c}{\textbf{Sep}} &
   \multicolumn{1}{c}{\textbf{PA}} &
      \multicolumn{1}{c}{\textbf{M$_{2}$}} &
    \multicolumn{1}{c}{\textbf{Sig.}} \\[0.2ex] 
   \multicolumn{1}{c}{$\mathbf{ }$} &
   \multicolumn{1}{c}{$\mathbf{(2400000 +) }$} &
   \multicolumn{1}{c}{\textbf{[mag]}} &
   \multicolumn{1}{c}{$\mathbf{ [mas]}$} &
   \multicolumn{1}{c}{$\mathbf{[au] }$} &
   \multicolumn{1}{c}{$\mathbf{[deg.] }$} &
   \multicolumn{1}{c}{\textbf{$\mathbf{ [ \,\Msun ]}$}} \\[-0.4ex]\hline \hline 
\multicolumn{7}{c}{}
\\[-3.2ex]

J04210934+2750368& 57001.04 &0.5&  19.28 $\pm$ 0.69 & 2.70 $\pm$ 0.09  & 309.60 $\pm$ 3.60&0.09&10.25\\
" & " & 1.0 &  19.01 $\pm$ 0.75 &2.66 $\pm$ 0.10& 307.68 $\pm$ 3.92&0.07&9.87\\
"& " &2.0 &  21.78 $\pm$ 1.32 &3.05 $\pm$ 0.18& 306.30 $\pm$ 4.63&0.04&8.54\\
"& " &3.0 &  26.40 $\pm$ 3.13 & 3.70 $\pm$ 0.94&306.89 $\pm$ 6.78&0.02&6.74\\
 \hline 

\end{tabular}
\end{minipage}
\end{table*}

Our fits to closure phases showed detections of 7 new candidate companions. As shown in section \ref{Sec:datanalysis}, we computed the Bayes' Factors for every object in the sample. Following the interpretation of \citet{Jeffreys1961}, we have found very strong Bayes' factors as an indication of the presence of candidate companions for LRL 31, V$*$V X-ray 6, WSB 12 and 2MASS J16335560-2442049 (see Table \ref{table:bayes}). For WSB 40, 2MASS J04210934+2750368 and 2MASS J16315473-2503238, we obtained moderate Bayes' factors ($>$10) from closure-phase alone, and considered these sources in more detail. In all cases, a visibility amplitude signal was found that was consistent with the best closure phase solution. In the case of WSB~40, the use of visibility amplitudes resulted in a clear solution with little degeneracy, as reported in Table \ref{Table:parameters}. We assigned a high ($>$300) Bayes' factor to this object in Table \ref{table:bayes}. In the case of 2MASS J04210934+2750368, contrast ratio and separation were highly degenerate, so we list possible statistically significant solutions at plausible contrast ratios as shown in Table \ref{table:reported}. For reported 2MASS J16315473-2503238, no binary solution was statistically significant \citep[taken as 6$\sigma$, e.g.][]{Kraus2016}, and there was only 1 epoch on the target under variable Laser Guide Star conditions. For these reasons, we do not report a binary solution, but note that a binary companion was confirmed as a Double-Line Spectroscopic binary (SB2) star composed of a K7 and a K9, with a semi-major axis of $<$ 0.6 au by \citep{Kohn2016} and therefore, we assigned a high ($>$300) Bayes' factor to this object in our statistical analysis (Table \ref{table:bayes}).

An interesting case is 2MASS J16335560-2442049, which was initially presented as a giant planet-forming candidate based on the morphology of its SED, large disk mass and modest accretion rate \citep{Cieza2010, Orellana2012}. Our $\chi^{2}$ minimization detected a secondary star located  at $\sim$ 3.25 $\pm$ 0.07 au and using \citet{Baraffe2015} models and $\Delta$K mag, the stellar mass would correspond to $\sim$ 0.61 $M_{\odot}$. However, the interpretation of a single-epoch for this object has to be taken with caution because of the high inclination ($\sim$ 50 $deg$) of its disk \citep{Cieza2012_tran32}, and the known degeneracy between the contrast ratio and small angular separations in the NRM data \citep{Pravdo2006}. Recently, \citet{Cieza2013} demonstrated that the starlight scattered off the inner edge of the FL Tau disk could mimic the presence of a faint companion, which might be the case of 2MASS J16335560-2442049. Further observing epochs are needed to establish the physical origin of the non-zero closure phases found in our analysis. Other cases of new binary systems are WSB 40  and WSB 12, which were below the detection limits of \citet{Cheetham2015}. Our careful reduction process and fits to closure phases resolved companions at $\sim$ 2.22 $\pm$ 0.12 au with a mass of $\sim$ 0.75 $M_{\odot}$ for WSB 40 and at $\sim$ 2.60 $\pm$ 0.10 au with a mass of $\sim$ 0.75 $M_{\odot}$ for WSB 12.

\subsection{Stars without a companion in the Inner Radii}

With a Bayes' factor of $<$ 0.1, the NRM data analysis did not detect binary stars with angular separations ranging from  $\sim \onethird$ to $ \twothirds$ of the $r_{d}$ for LRL 21, LRL 67, IRAS04125+2902, EM* SR 21A, DoAr 44, 2MASS J16262367-2443138, 2MASS J16273901-2358187, WSB 63, EM* SR 24S and V*V852 Oph (Table \ref{table:bayes}).  Detection limits for these objects are listed in Table \ref{table:nondetections}. The absence of a binary companion implies that the inner region mainly is being dispersed by an internal process that determines the lifetime of the disk. To date, the different mechanisms proposed to explain the inner holes of these disks do not accurately predict the observed features of TDs, and produced theoretically distinct properties of the TDs \citep[e.g.][]{Alexander2007}. Despite all the efforts to develop a unique explanation of the evolution of the disk and its transition phase from a class II to III, \textit{photoevaporation} and \textit{planetary formation}, and its counterpart processes such as dust filtration and grain growth, seem to be the most efficient mechanisms to disperse the disk from the inside out.

 \subsection{Unresolved Transitional Disks}

 In Table~\ref{table:bayes}, LRL 237, LRL 97, LRL 58 and 2MASS J16250692-2350502 are the objects with Bayes' factors of $\sim$ 1, meaning that our NRM observations were not able to rule out or confirm companions for those objects, where the inner radii estimations fall inside our detection limit of 25 mas. These objects have M1 or later spectral types, meaning that they are at the low mass end of our sample. They also have relatively low accretion rates, typical of lower mass objects  \citep[][$\lesssim$ 10$^{-8}$  $M_{\odot}$yr$^{-1}$]{Herczeg2008} . If their lack of a NIR excess is due to clearing by a binary companion, such a companion can only be discovered by multi-epoch radial velocity monitoring. We summarise our detection limits in Table \ref{table:nondetections}.
 
 \section{Statistics}
\label{sec:stats}

After estimating the stellar properties and computing the Bayes' factor $\Upsilon_{B,S}=\frac{P(D|B)}{P(D|S)}$ for every target, we proceed to estimate the frequency of binary stars producing the ``transitional disk SEDs".

Continuing with Bayesian statistics, we assigned prior probabilities to each population, where the prior information for a binary or single system in TDs are equally probable. A non-informative probability distribution for the frequency of binary or single systems is a parameter representing two unique scenarios, binary or non-binary. Thus, $P(B)=\gamma$ represents the probability of binarity and $P(S)=(1-\gamma$) not binarity, where  the paramenter $\gamma$ $\in [0,1]$  and is sampled by the family of \textit{beta distributions} \footnote{Beta distribution has parameters $\alpha,  \beta=1/2$ to ensure a probability equal to unity and events equally probable.} \citep{Glickman2007}.

 The Jeffreys prior for the sampling distribution that provides uniform probability to both events B and S is represented by:

\begin{equation}
P(\gamma )\propto \frac{1}{\sqrt{\gamma (1-\gamma )}}\propto \frac{1}{\sqrt{P(B) P(S) }}.
\end{equation}

Taking into account that the current work is pioneering in the search for close binary companions ($>$ 40 mas) or confirming a single star in the inner region of TDs, no information was available from inside the inner region of these TDs. Therefore, we have to assume that the data have arisen from one of two systems and being equally probable. Thus, a prior distribution $\gamma$ with the form that represents the best prior state of knowledge can be modified with the observed data as follows:

\begin{equation}
P(\gamma |D)\propto  P(\gamma ) P(D |\gamma)\propto  P(\gamma )  [\gamma (\Upsilon _{B,S}) + (1-\gamma )]
\end{equation}

where $\Upsilon_{B,S}$ is the computed Bayes' factor of every TD.
Considering all TDs in our sample and additionally TDs from previous studies, the posterior density function of the frequency of binary systems responsible of the observed SEDs is:

\begin{equation}
P(\gamma |D)\propto  P(\gamma ) \prod_{i=1}^{i} [\gamma (\Upsilon _{B,S})_{i} + (1-\gamma )]
\label{eq:finalposterior}
\end{equation}

with the index $i$ representing the number of objects included in the modification of $\gamma$.

\subsection{Binary Frequency in TDs}

\begin{figure}[!h]
 \centering
 
    \includegraphics[width=0.5\textwidth]{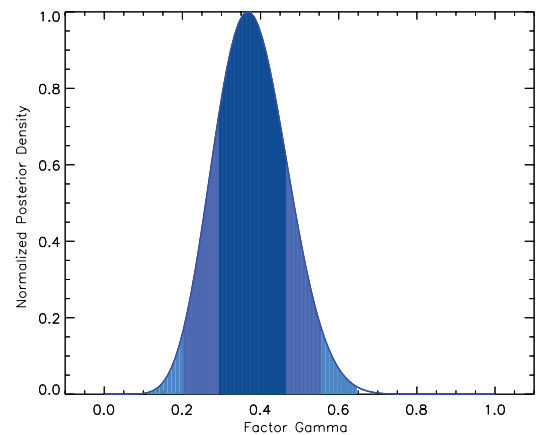}
 \caption{ The fraction of TD in our sample consistent with a binary system being the main mechanism causing the characteristic SED is 0.38 $\pm$ 0.09.}
 \label{fig:posterior}

\end{figure}

Binary and single objects were identified in our sample of TDs with Bayes' factors $ \Upsilon$ $<$ 0.1, $\Upsilon >$ 300 and  $\Upsilon \simeq$ 1.0, which are likely single stars, binary stars or unresolved systems, respectively, except for 2MASS J04210934+2750368 with a Bayes' factor of 24 (Table \ref{table:bayes}); we proceeded to include objects that were already classified as TDs and eventually, were characterized as binary or single stars\footnote{For the special cases of 2MASS J16315473-2503238 and previously studied, FL Cha, we opted to include them as binary objects in the statistical analysis.}. For these binary and single objects a Bayes' factor of 300 and 0.001 were used in the Bayesian statistical analysis, respectively.  These objects are summarised in Table \ref{table:extra}.
After computing the posterior probability, see Equation ~\ref{eq:finalposterior}, our uniform prior has been modified to the posterior probability of 0.38 $\pm$ 0.09 and shown in Figure ~\ref{fig:posterior}.

\restylefloat{figure}

\begin{table*}
 \centering
 \begin{minipage}{130mm}
  \caption{Stellar Properties of Other Known TDs}
 \label{table:extra}
  \begin{tabular}{lcccccccc}
\hline \hline \\[-3ex]
    \multicolumn{1}{c}{\textbf{Object}} &
    \multicolumn{1}{c}{\textbf{R.A. (J2000)}} &
    \multicolumn{1}{c}{\textbf{Dec. (J2000)}}&
    \multicolumn{1}{c}{\textbf{Type}} &
    \multicolumn{1}{c}{\textbf{Spec. Type}} &
    \multicolumn{1}{c}{\textbf{Accretor}} &
    \multicolumn{1}{c}{\textbf{R$_{cav.}$}} &
    \multicolumn{1}{c}{\textbf{Reference}} \\[0.2ex] 
     \multicolumn{1}{c}{\textbf{}} &
      \multicolumn{1}{c}{\textbf{}} &
       \multicolumn{1}{c}{\textbf{}} &
        \multicolumn{1}{c}{\textbf{}} &
     \multicolumn{1}{c}{\textbf{}} &
      \multicolumn{1}{c}{\textbf{}} &
      \multicolumn{1}{c}{\textbf{[au]}} &
      \multicolumn{1}{c}{\textbf{}}   \\[-0.4ex]\hline \hline 
   
\multicolumn{8}{c}{}
\\[-3.2ex]

Coku Tauri/4&04 41 16.808&+28 40 00.07&Binary&M1.5&n&10&1, 2,11\\
LKCa15  & 04 39 17.796 &+22 21 03.48&Single&K5v&y&50&1,3,5,9 \\  
DM Tau &04 33 48.73 & +18 10 10.0&Single& M1&y&19&1,4,5,9\\
GM Aur&04 55 10.983 &+30 21 59.54&Single&K5&y&28&1,4,5,9\\
UX Tau A&04 30 03.988 &+18 13 49.61&Single&G8&y&25&1,4,5,9\\
RY Tau&04 21 57.41 &+28 26 35.57&Single&K1&y&18&1,5,10\\
CS Cha & 11 02 24.912&-77 33 35.72&Binary&K6&y&38&1,6,10\\
T Cha & 11 57 13.550& -79 21 31.54&Single&K0&y&40&1,6,12\\
FL Cha &11 08 39.051& -77 16 04.24&Binary&K8&y &8.3&1,7\\
TW Hydrae&11 01 51.907 &-34 42 17.03&Single&K6&y&41&1,8,13\\
Haro 1-16 & 16 31 33.46 & -24 27 37.3 & Single & K3 &y&36&1, 14, 15\\
 \hline 
 \end{tabular}
 
Reference: (1) 2MASS All-Sky Point Source Catalog, (2) \citet{Ireland2008},  (3) \citet{Kraus2012}, (4)\citet{Huelamo2011}, (5) \citet{Pott2010}, 
(6) \citet{Guenther2007}, (7) \citet{Cieza2013}, (8) \citet{Rapson2015}, (9) \citet{Andrews2011}, (10) \citet{Espaillat2011}, (11) \citet{Dalessio2005}, 
(12) \citet{Huelamo2015}, (13) \citet{ Nomura2016}, (14) \citet{Bouvier1992}, (12) \citet{Cheetham2015}

\end{minipage}
\end{table*}

\section{Discussion}

In our combined sample consisting of 31 objects, including 11 TDs and CDs with known multiplicity from the literature, and excluding 3 wide binaries, we find that a fraction of 0.38 $\pm$ 0.09 of the SEDs are being produced by the flux emission of a binary star $+$ disk instead of a single star $+$ disk. This means that the remaining SEDs with low NIR and MIR excesses observed to date are the result of the dispersion of the primordial material due to another internal mechanism. 
Our binary detections inside the fitted disk wall inner radii do not necessarily have projected separations between $\onethird$ and $\onehalf$ of the inner radii, which is the expected semi-major axis range for a binary to cause the truncation of the disk. However, all detections lie within $\onehalf$ of our calculated inner disk radii, consistent with projection effects.

Given the criteria applied to select our sample and following the standards for disk classification, we emphasize that these objects should be treated as CDs that possibly are in a \textit{transitional phase}, and no longer treat them as TDs with a single star. Originally, the SEDs of these objects were described assuming only one object in the interior of the disk and using detailed disk models to fit the excess continua \citep[e.g.][]{Espaillat2012}. As demonstrated by this work, there is a significant fraction of these SEDs which were mis-classified. However, as seen in Figure \ref{fig:imagessed}, the CD SEDs of LRL 31,  V410 X-ray 6, WSB 12, WSB 40 , 2MASS J16335560-2442049, 2MASS J04210934+2750368 and
2MASS J16315473-2503238 are indistinguishable from TDs. Although, to date the resemblance between CD SEDs and TD SEDs is well established \citep[e.g.][]{Ireland2008}, unfortunately we could not set an observational constraint such as accretion rate or flux emission in our sample. For example, the SEDs of V410 X-ray 6 and 2MASS J16335560-2442049 bear a resemblance to the large MIR emission and zero NIR excess detected in the binary Coku Tau/4  \citep{Dalessio2005, Ireland2008}, while the other objects show a more similar SED to a typical TD SED. On the other hand, for those objects shown in Table \ref{table:bayes} with Bayes' factors $\simeq$ 1, that due to resolution limitations we were not able to confirm or rule out their binarity, multi-epoch RV monitoring observations are needed \citep[e.g.][]{Kohn2016}, because there might be more binary objects dispersing the inner region of the disk efficiently. 

We have also detected 4 new binary systems with the location of the secondary component outside the inner region of the disk. Interestingly, these systems produced SEDs characteristic of the TDs and are low accretors (Table~\ref{table:nosample}). We have proposed that those SEDs composed of a low-mass binary star with one of its components orbiting outside the inner radius of the disk, might have its more ``evolved'' disk orbiting the sub-stellar companion, instead of the primary component. Although, it is also plausible that the primary component has a circumstellar disk that is being dispersed by the close sub-stellar companion. Previously, \citet{Harris2012} performed a high angular resolution millimeter-wave dust continuum imaging survey of circumstellar material associated with the individual components of multiple star systems in the Taurus$-$Auriga young cluster. They found that the presence of a close stellar companion ($<$ 30 au ) impacts disk properties, producing a disk mass depletion with a factor of $\sim$ 25. In the case of the LRL 72, LRL 182, LRL 213 and LRL 135 systems, a faster dispersion of the disk by the presence of the stellar companion located at $\leq$ 20 au could influence the initial conditions for the formation of planets and prevent the first steps of this evolutionary process (e.g. dust settling and grain growth).

\subsection{Physical Sources of Typical TD SEDs}

Planetary formation could potentially explain the estimated inner optically thick disk radii for these objects and therefore, the peculiar shape or decreased flux observed in the NIR/MIR SEDs of these TDs. Depending on the inner hole size, the gap could be cleared up by single or multiple planets orbiting this region \citep{ Lubow1999, Rice2006, Dodson2011}. In the context of planet disk interaction, and as a consequence of a massive planet clearing out the inner region of the disk, a local pressure bump is created at the inner edge of the outer disk. In the last decade, this local pressure bump was proposed to act as a filter at the outer edge of a disk gap, filtering particles of size $\gtrsim$ 10 $\micron$ and impeding the drift inward of them \citep{Rice2006a}. As a result of this \textit{dust filtration}, the disk profile is shown with an abrupt discontinuity in its dust radial profile and at the same time permits the presence of small particles closer to the central star ($\lesssim$ 10 $\micron$) \citep[e.g][]{Garufi2013}. Thus, this optically thin dust might be responsible of the weak NIR/MIR excess present in TD SEDs. In addition, inside this cavity coupling between $\micron$ size dust grains and gas  is expected \citep{Garufi2013}, while the location to pile-up the dust at a sub- to millimeter scale in a pressure maximum, leads to different locations of gap edges for gas and ``bigger" dust particles \citep{Pinilla2012}. In our approach to estimate inner cavities, we consider the location of particles of $\sim$ 0.1 $\micron$ that might coincide with the gaseous cavity, ingredients necessary to explain the detected accretion rates in our sample of TDs.

Most of our TDs show accretion rates ranging from 10$^{-8}$ to 10$^{-10} $ $M_{\odot}$/yr and, although these accreting TDs are also ideal targets to test the role of some photoevaporation models \citep[e.g.][]{Alexander2006a, Alexander2006}, there are other missing pieces to the puzzle such as disk mass measurements needed to obtain a complete picture of this transitional phase. Therefore, the observed SEDs of TDs with the presence of a single star might be subject to a dominating internal mechanism and the amount of mass in the disk. Thus, in order to distinguish the dominating dispersal mechanism producing the inner holes in the disks, a follow-up program of millimeter observations of the TDs is required to be able to estimate the disk mass of these objects. Nevertheless, the inner region of these TDs could be depleted by a combination of two or more mechanisms that dominate at different distances from the central star and timescales dictated by the initial physical conditions.

\subsection{Single vs. Binary Stars: Hosting Planetary Formation}

At first glance, it is tempting to suggest that single stars have a higher probability of hosting the formation of planetary systems than close binary systems. However, \citet{Pascucci2008} studied the first steps of planetary formation in single and binary systems with projected separations between $\sim$ 10 and 450 au and they found no statistical significant difference in the degree of dust settling and grain growth of those systems, indicating that expected differences in the exoplanet properties arise in the later stages of their formation and/or migration \citep[e.g.][]{Kley2000, Kley2007}. Our close binary companions are detected at angular separations between 2$-$10 au; these small angular separations might affect the initial conditions for the formation of planets in the inner region of the circumbinary disks. This is mainly due to the modification of the binary eccentricity and excitation of density waves generated by the resonant interactions of the binaries with the disk, which remove primordial material \citep{Lubow2000}. Based on these assumptions, the ``weak'' excess from the circumstellar material in the SEDs of the CDs, increased by the secondary flux radiation, could point out a lower probability for the formation of a planet in radii of around $a \leq 10$ au in very close binary stars. On the other hand, single stars are more probable to host forming planets at inner radii  around $<$ 10 au than close binary stars, where actually most of the planet formation might take place.

Because the time available to form any planet(s) in a circumstellar disk might vary depending on the initial conditions and the evolution of the disk, it is necessary in future surveys to characterize the distribution of disk masses in CDs with close binary and single stars, that together with the accretion rates will establish the physical parameters constraining where and when planets form in those systems. Additionally, accretion rates have been used to estimate the dissipation of the primordial disks once accretion stops; however, we did not find any trend in $\dot{M}_{*}$ or difference between close binary and single stars in our sample that helps us to constrain the timescales of these systems.

\section{Summary}

Using infrared NRM interferometry taken with the Keck II telescope, we have observed a sample of 24 TDs located in the Taurus-Auriga, IC-348 and Ophiuchus star-forming regions. We implemented a new method of completeness correction for our detection limits, which combines randomly sampled binary orbits and Bayesian inference. With high confidence levels of 99$\%$, a total of 7 close binary candidates have been detected orbiting the inner radii of the TDs, and likely being the main mechanism responsible for the dispersion inside out the inner disk. Also, we found four binary companions orbiting outside the inner radii of their TDs  and we have suggested that the unusual SEDs of these systems are due to a disk orbiting a substellar secondary companion, producing similar SEDs as the single and/or close binary stars surrounded by a more ``evolved" disk or weak disk.

Including 11 known TDs from the literature and whose binarity was already confirmed or ruled out, we have a total of 31 TDs that are part of our Bayesian analysis (Section \ref{sec:stats}). Updating a uniform prior distribution, we obtained a significant fraction of 0.38 $\pm$ 0.09 objects with TD SEDs that are actually CDs. This fraction represents the unusual SEDs with a lack of excess in the NIR and/or MIR being produced by the flux emission of a close-binary companion and a disk. This fraction must be taken into consideration for future surveys and studies of these transitional objects in order to decode the disk evolution process and the timescales of close binary and single stars, separately.  The remaining SEDs are being produced by a single system and a disk in a transition phase, where the main cause of dispersion could be any other internal mechanism such as photoevaporation, grain growth and/or planet disk interactions.

\section{Acknowledgements}

We thank the referee for her/his comments and suggestions, which helped in the improvement of the paper. M.I was suported by the Australian Research Council's Future Fellowship scheme (FT130100235). L.A.C. acknowledges support from the Millennium Science Initiative (Chilean Ministry of Economy),  through grant ÒNucleus  RC130007Ó. L.A.C. was also supported by CONICYT-FONDECYT grant number 1140109.

\begin{appendix}

\section{Distance to IC348 Region}

The  distance to the Perseus molecular cloud has been measured in a wide range between 220 - 380 pc  \citep{Harris1954, Herbig1983, Cernis1993, Scholz1999, deZeeuw1999, Hirota2008, Hirota2010}, and choosing the most appropriate measurement must be taken with caution because of the influence on other observational estimations, such as age and luminosity of the targets. For our purposes, we base our decision to adopt a distance to IC348 on the astrometric observations of H$_{2}$O maser sources by \citet{Hirota2010}. They used the VERA long-baseline array to estimate a distance to SVS 13 in the NGC 1333 cluster of 235 $\pm$ 18 pc and  a distance of 232 $\pm$ 18 pc to L1448. In addition, \citet{Sun2006} mapped the Perseus molecular cloud complex simultaneously in $^{13}$CO(J=2-1) and $^{12}$CO(J=2-1) using the KOSMA 3 m submillimeter telescope. They found a  dynamical connection with a velocity gradient between NGC 1333 at $\sim$7 km $s^{-1}$,  L 1448 at $\sim$8 km $s^{-1}$ and IC 348 at $\sim$9 km $s^{-1}$ within a diameter of 20 pc of extension. Considering that the Perseus molecular cloud  \citep[containing IC348, NGC1333, L1448, L1445;][]{Hirota2010} has a full angular extent of only $\pm$3$^{^{\circ}}$ from its centre, it is expected to have a characteristic distance range of 5\%.

\begin{table}
 \centering
 \begin{minipage}{38mm}
  \caption{IC348 1 Properties}
 \label{table:ic3481}
  \begin{tabular}{lc}
\hline \hline \\[-3ex]
    \multicolumn{1}{c}{\textbf{Property}} &    
    \multicolumn{1}{c}{\textbf{Value}} \\[-0.2ex] 
     \multicolumn{1}{c}{\textbf{}} &
      \multicolumn{1}{c}{\textbf{}}   \\[-2.4ex]\hline \hline 
R.A. (J2000)\footnotemark[1]&03:44:56.15\\
Dec. (J2000)\footnotemark[1] & +32:09:15.5 \\
Spec. Type&B5\\
$L_{*}$ ($L_{\odot}$)\footnotemark[2] & 324.8\\  
$A_{J}$\footnotemark[3]&0.55\\
$T_{*}$ (K)& 15400\\
$M_{*}$ ($M_{\odot}$)&4.6\\
Dist. (pc)& 220\\
 \hline 
 \end{tabular}
Reference: (1) 2MASS All-Sky Point Source Catalog, (2) Luminosity estimated following the method from \citet{Kenyon1995}, (3) A$_{J}$ estimated as described in Section \ref{sec:stellarparameter}, (4) Spectral type and stellar temperature were adopted from \citet{Luhman2003}.
\end{minipage}
\end{table}

 \begin{figure}
 \centering
     \includegraphics[width=0.53\textwidth]{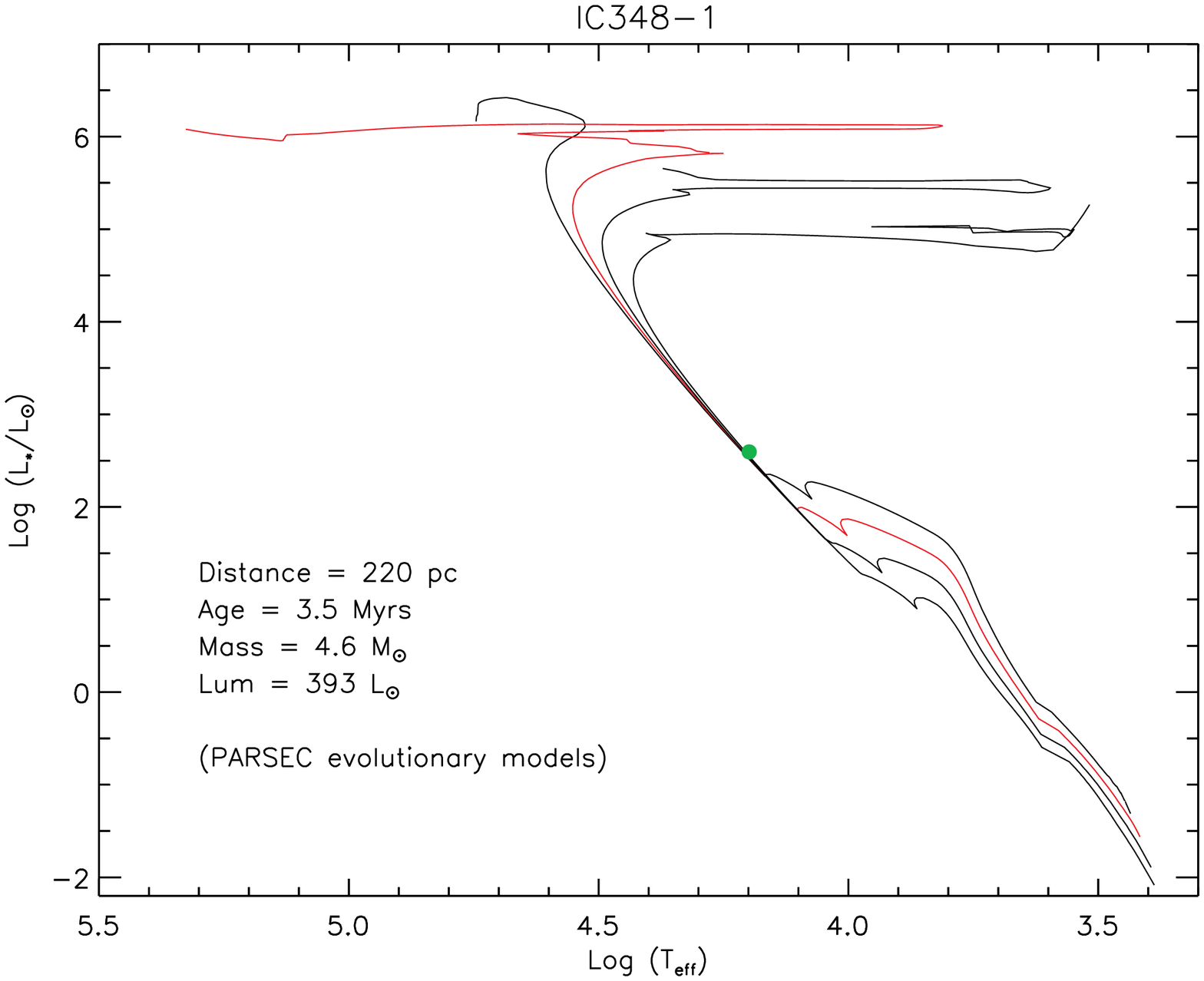}
             \caption{PARSEC evolutionary models for young stars \citep{Bressan2012}. Black Solid lines in descending order are 2, 3.5, 6 and 10 Myrs isochrones. Red solid line corresponds to the 3.5 Myrs isochrone used to characterised IC348$-$1 and then, estimate a distance to the IC348 star-forming region.}
  \label{fig:padova} 
 \end{figure}

Here, we use the most luminous member of the cluster, IC 348 LRL 1, composed of two stars of similar brightness with spectral types B5V to adopt an independent distance to IC348 objects. IC 348 LRL 1 is a binary system with an angular separation of $0.47"$ and P.A. of 17.9$^{^{\circ}}$ \citep{Alzner1998} and as the only member of B spectral type in the region, it can be used to estimate a distance to it \citep{Luhman1998}. At early ages, $\leq $ 10 Myrs, the bolometric luminosity of these B spectral type massive stars do not vary significantly, giving an independence of the isochrone used to describe the target. This allows a spectral type conversion into effective temperature without constraints on the age of the system. Figure \ref{fig:padova} presents the isochrones of 2, 3.5, 6 and 10 Myrs, as they are the most representative to describe young members of IC348. Clearly, the isochrones used to predict the stellar parameters for  IC 348 LRL 1 are mostly invariant at this early stage of the stellar evolution.

We derived the stellar properties of IC 348 LRL 1 based on the spectral type and a conversion to the stellar temperature, see Table \ref{table:ic3481}. According to the PARSEC evolutionary models \citep{Bressan2012}, a B5V object with a stellar mass of 4.2 M$_{\odot }$ should be located at a distance of 220 $\pm$ 10 pc, see Figure \ref{fig:padova}. Thus, a stellar luminosity of 393 L$_{\odot }$ is calculated with the dereddened J-band photometry method from  \citet{Kenyon1995} and A$_{J}$  extinction was estimated as explained in Section \ref{sec:stellarparameter}.

\end{appendix}

\clearpage
\bibliographystyle{mn2e}
\bibliography{biblio}

\begin{thebibliography}{110}
\expandafter\ifx\csname natexlab\endcsname\relax\def\natexlab#1{#1}\fi

\bibitem[{{Alexander} \& {Armitage}(2007)}]{Alexander2007}
{Alexander} R.~D., {Armitage} P.~J., 2007, \mnras, 375, 500

\bibitem[{{Alexander}, {Clarke} \& {Pringle}(2006{\natexlab{a}}){Alexander},
  {Clarke}, \& {Pringle}}]{Alexander2006a}
{Alexander} R.~D., {Clarke} C.~J., {Pringle} J.~E., 2006{\natexlab{a}}, \mnras,
  369, 216

\bibitem[{{Alexander}, {Clarke} \& {Pringle}(2006{\natexlab{b}}){Alexander},
  {Clarke}, \& {Pringle}}]{Alexander2006}
{Alexander} R.~D., {Clarke} C.~J., {Pringle} J.~E., 2006{\natexlab{b}}, \mnras,
  369, 229

\bibitem[{{Allard}(2014)}]{Allard2014}
{Allard} F., 2014, in IAU Symposium, Vol. 299, IAU Symposium, {Booth} M.,
  {Matthews} B.~C., {Graham} J.~R., eds., pp. 271--272

\bibitem[{{Alzner}(1998)}]{Alzner1998}
{Alzner} A., 1998, \aaps, 132, 237

\bibitem[{{Ambartsumian}(1937)}]{Ambartsumian1937}
{Ambartsumian} V.~A., 1937, Astron. Zh., 14, 207

\bibitem[{{Andrews} {et~al}\mbox{.}(2011){Andrews}, {Wilner}, {Espaillat},
  {Hughes}, {Dullemond}, {McClure}, {Qi}, \& {Brown}}]{Andrews2011}
{Andrews} S.~M., {Wilner} D.~J., {Espaillat} C., {Hughes} A.~M., {Dullemond}
  C.~P., {McClure} M.~K., {Qi} C., {Brown} J.~M., 2011, \apj, 732, 42

\bibitem[{{Artymowicz} \& {Lubow}(1994)}]{Artymowicz1994}
{Artymowicz} P., {Lubow} S.~H., 1994, \apj, 421, 651

\bibitem[{{Baraffe} {et~al}\mbox{.}(2015){Baraffe}, {Homeier}, {Allard}, \&
  {Chabrier}}]{Baraffe2015}
{Baraffe} I., {Homeier} D., {Allard} F., {Chabrier} G., 2015, \aap, 577, A42

\bibitem[{{Bayo} {et~al}\mbox{.}(2008){Bayo}, {Rodrigo}, {Barrado Y
  Navascu{\'e}s}, {Solano}, {Guti{\'e}rrez}, {Morales-Calder{\'o}n}, \&
  {Allard}}]{Bayo2008}
{Bayo} A., {Rodrigo} C., {Barrado Y Navascu{\'e}s} D., {Solano} E.,
  {Guti{\'e}rrez} R., {Morales-Calder{\'o}n} M., {Allard} F., 2008, \aap, 492,
  277

\bibitem[{{Bouvier} \& {Appenzeller}(1992)}]{Bouvier1992}
{Bouvier} J., {Appenzeller} I., 1992, \aaps, 92, 481

\bibitem[{{Bressan} {et~al}\mbox{.}(2012){Bressan}, {Marigo}, {Girardi},
  {Salasnich}, {Dal Cero}, {Rubele}, \& {Nanni}}]{Bressan2012}
{Bressan} A., {Marigo} P., {Girardi} L., {Salasnich} B., {Dal Cero} C.,
  {Rubele} S., {Nanni} A., 2012, \mnras, 427, 127

\bibitem[{{Canovas} {et~al}\mbox{.}(2016){Canovas}, {Caceres}, {Schreiber},
  {Hardy}, {Cieza}, {M{\'e}nard}, \& {Hales}}]{Canovas2016}
{Canovas} H., {Caceres} C., {Schreiber} M.~R., {Hardy} A., {Cieza} L.,
  {M{\'e}nard} F., {Hales} A., 2016, ArXiv e-prints

\bibitem[{{Cernis}(1993)}]{Cernis1993}
{Cernis} K., 1993, Baltic Astronomy, 2, 214

\bibitem[{{Cheetham} {et~al}\mbox{.}(2015){Cheetham}, {Kraus}, {Ireland},
  {Cieza}, {Rizzuto}, \& {Tuthill}}]{Cheetham2015}
{Cheetham} A.~C., {Kraus} A.~L., {Ireland} M.~J., {Cieza} L., {Rizzuto} A.~C.,
  {Tuthill} P.~G., 2015, \apj, 813, 83

\bibitem[{{Chiang} \& {Murray-Clay}(2007)}]{ChiangMurray2007}
{Chiang} E., {Murray-Clay} R., 2007, Nature Physics, 3, 604

\bibitem[{{Cieza} {et~al}\mbox{.}(2013){Cieza}, {Lacour}, {Schreiber},
  {Casassus}, {Jord{\'a}n}, {Mathews}, {C{\'a}novas}, {M{\'e}nard}, {Kraus},
  {P{\'e}rez}, {Tuthill}, \& {Ireland}}]{Cieza2013}
{Cieza} L.~A. {et~al.}, 2013, \apjl, 762, L12

\bibitem[{{Cieza} {et~al}\mbox{.}(2012{\natexlab{a}}){Cieza}, {Mathews},
  {Williams}, {M{\'e}nard}, {Kraus}, {Schreiber}, {Romero}, {Orellana}, \&
  {Ireland}}]{Cieza2012_tran32}
{Cieza} L.~A. {et~al.}, 2012{\natexlab{a}}, \apj, 752, 75

\bibitem[{{Cieza} {et~al}\mbox{.}(2010){Cieza}, {Schreiber}, {Romero}, {Mora},
  {Merin}, {Swift}, {Orellana}, {Williams}, {Harvey}, \& {Evans}}]{Cieza2010}
{Cieza} L.~A. {et~al.}, 2010, \apj, 712, 925

\bibitem[{{Cieza} {et~al}\mbox{.}(2012{\natexlab{b}}){Cieza}, {Schreiber},
  {Romero}, {Williams}, {Rebassa-Mansergas}, \& {Mer{\'{\i}}n}}]{Cieza2012}
{Cieza} L.~A., {Schreiber} M.~R., {Romero} G.~A., {Williams} J.~P.,
  {Rebassa-Mansergas} A., {Mer{\'{\i}}n} B., 2012{\natexlab{b}}, \apj, 750, 157

\bibitem[{{Clarke}, {Gendrin} \& {Sotomayor}(2001){Clarke}, {Gendrin}, \&
  {Sotomayor}}]{Clarke2001}
{Clarke} C.~J., {Gendrin} A., {Sotomayor} M., 2001, \mnras, 328, 485

\bibitem[{{Currie} \& {Kenyon}(2009)}]{Currie2009}
{Currie} T., {Kenyon} S.~J., 2009, \aj, 138, 703

\bibitem[{{Cutri} {et~al}\mbox{.}(2003){Cutri}, {Skrutskie}, {van Dyk},
  {Beichman}, {Carpenter}, {Chester}, {Cambresy}, {Evans}, {Fowler}, {Gizis},
  {Howard}, {Huchra}, {Jarrett}, {Kopan}, {Kirkpatrick}, {Light}, {Marsh},
  {McCallon}, {Schneider}, {Stiening}, {Sykes}, {Weinberg}, {Wheaton},
  {Wheelock}, \& {Zacarias}}]{Cutri2003}
{Cutri} R.~M. {et~al.}, 2003, VizieR Online Data Catalog, 2246

\bibitem[{{D'Alessio} {et~al}\mbox{.}(2005){D'Alessio}, {Hartmann}, {Calvet},
  {Franco-Hern{\'a}ndez}, {Forrest}, {Sargent}, {Furlan}, {Uchida}, {Green},
  {Watson}, {Chen}, {Kemper}, {Sloan}, \& {Najita}}]{Dalessio2005}
{D'Alessio} P. {et~al.}, 2005, \apj, 621, 461

\bibitem[{{de Zeeuw} {et~al}\mbox{.}(1999){de Zeeuw}, {Hoogerwerf}, {de
  Bruijne}, {Brown}, \& {Blaauw}}]{deZeeuw1999}
{de Zeeuw} P.~T., {Hoogerwerf} R., {de Bruijne} J.~H.~J., {Brown} A.~G.~A.,
  {Blaauw} A., 1999, \aj, 117, 354

\bibitem[{{Dodson-Robinson} \& {Salyk}(2011)}]{Dodson2011}
{Dodson-Robinson} S.~E., {Salyk} C., 2011, \apj, 738, 131

\bibitem[{{Duch{\^e}ne} {et~al}\mbox{.}(2003){Duch{\^e}ne}, {Ghez}, {McCabe},
  \& {Weinberger}}]{Duchene2003}
{Duch{\^e}ne} G., {Ghez} A.~M., {McCabe} C., {Weinberger} A.~J., 2003, \apj,
  592, 288

\bibitem[{{Dullemond}, {Dominik} \& {Natta}(2001){Dullemond}, {Dominik}, \&
  {Natta}}]{Dullemond2001}
{Dullemond} C.~P., {Dominik} C., {Natta} A., 2001, \apj, 560, 957

\bibitem[{{Dullemond} \& {Monnier}(2010)}]{Dullemond2010}
{Dullemond} C.~P., {Monnier} J.~D., 2010, \araa, 48, 205

\bibitem[{{Duquennoy} \& {Mayor}(1991)}]{Duquennoy1991}
{Duquennoy} A., {Mayor} M., 1991, \aap, 248, 485

\bibitem[{{Espaillat} {et~al}\mbox{.}(2015){Espaillat}, {Andrews}, {Powell},
  {Feldman}, {Qi}, {Wilner}, \& {DAlessio}}]{Espaillat2015}
{Espaillat} C., {Andrews} S., {Powell} D., {Feldman} D., {Qi} C., {Wilner} D.,
  {DAlessio} P., 2015, \apj, 807, 156

\bibitem[{{Espaillat} {et~al}\mbox{.}(2007){Espaillat}, {Calvet}, {D'Alessio},
  {Hern{\'a}ndez}, {Qi}, {Hartmann}, {Furlan}, \& {Watson}}]{Espaillat2007}
{Espaillat} C., {Calvet} N., {D'Alessio} P., {Hern{\'a}ndez} J., {Qi} C.,
  {Hartmann} L., {Furlan} E., {Watson} D.~M., 2007, \apjl, 670, L135

\bibitem[{{Espaillat} {et~al}\mbox{.}(2010){Espaillat}, {D'Alessio},
  {Hern{\'a}ndez}, {Nagel}, {Luhman}, {Watson}, {Calvet}, {Muzerolle}, \&
  {McClure}}]{Espaillat2010}
{Espaillat} C. {et~al.}, 2010, \apj, 717, 441

\bibitem[{{Espaillat} {et~al}\mbox{.}(2011){Espaillat}, {Furlan}, {D'Alessio},
  {Sargent}, {Nagel}, {Calvet}, {Watson}, \& {Muzerolle}}]{Espaillat2011}
{Espaillat} C., {Furlan} E., {D'Alessio} P., {Sargent} B., {Nagel} E., {Calvet}
  N., {Watson} D.~M., {Muzerolle} J., 2011, \apj, 728, 49

\bibitem[{{Espaillat} {et~al}\mbox{.}(2012){Espaillat}, {Ingleby},
  {Hern{\'a}ndez}, {Furlan}, {D'Alessio}, {Calvet}, {Andrews}, {Muzerolle},
  {Qi}, \& {Wilner}}]{Espaillat2012}
{Espaillat} C. {et~al.}, 2012, \apj, 747, 103

\bibitem[{{Espaillat} {et~al}\mbox{.}(2014){Espaillat}, {Muzerolle}, {Najita},
  {Andrews}, {Zhu}, {Calvet}, {Kraus}, {Hashimoto}, {Kraus}, \&
  {D'Alessio}}]{Espaillat2014}
{Espaillat} C. {et~al.}, 2014, Protostars and Planets VI, 497

\bibitem[{{Evans} {et~al}\mbox{.}(2003){Evans}, {Allen}, {Blake}, {Boogert},
  {Bourke}, {Harvey}, {Kessler}, {Koerner}, {Lee}, {Mundy}, {Myers}, {Padgett},
  {Pontoppidan}, {Sargent}, {Stapelfeldt}, {van Dishoeck}, {Young}, \&
  {Young}}]{Evans2003}
{Evans}, II N.~J. {et~al.}, 2003, \pasp, 115, 965

\bibitem[{{Evans} {et~al}\mbox{.}(2009){Evans}, {Dunham}, {J{\o}rgensen},
  {Enoch}, {Mer{\'{\i}}n}, {van Dishoeck}, {Alcal{\'a}}, {Myers},
  {Stapelfeldt}, {Huard}, {Allen}, {Harvey}, {van Kempen}, {Blake}, {Koerner},
  {Mundy}, {Padgett}, \& {Sargent}}]{Evans2009a}
{Evans}, II N.~J. {et~al.}, 2009, \apjs, 181, 321

\bibitem[{{Flaherty} {et~al}\mbox{.}(2011){Flaherty}, {Muzerolle}, {Rieke},
  {Gutermuth}, {Balog}, {Herbst}, {Megeath}, \& {Kun}}]{Flaherty2011}
{Flaherty} K.~M., {Muzerolle} J., {Rieke} G., {Gutermuth} R., {Balog} Z.,
  {Herbst} W., {Megeath} S.~T., {Kun} M., 2011, \apj, 732, 83

\bibitem[{{Forrest} {et~al}\mbox{.}(2004){Forrest}, {Sargent}, {Furlan},
  {D'Alessio}, {Calvet}, {Hartmann}, {Uchida}, {Green}, {Watson}, {Chen},
  {Kemper}, {Keller}, {Sloan}, {Herter}, {Brandl}, {Houck}, {Barry}, {Hall},
  {Morris}, {Najita}, \& {Myers}}]{Forrest2004}
{Forrest} W.~J. {et~al.}, 2004, \apjs, 154, 443

\bibitem[{{Furlan} {et~al}\mbox{.}(2011){Furlan}, {Luhman}, {Espaillat},
  {D'Alessio}, {Adame}, {Manoj}, {Kim}, {Watson}, {Forrest}, {McClure},
  {Calvet}, {Sargent}, {Green}, \& {Fischer}}]{Furlan2011}
{Furlan} E. {et~al.}, 2011, \apjs, 195, 3

\bibitem[{{Garufi} {et~al}\mbox{.}(2013){Garufi}, {Quanz}, {Avenhaus},
  {Buenzli}, {Dominik}, {Meru}, {Meyer}, {Pinilla}, {Schmid}, \&
  {Wolf}}]{Garufi2013}
{Garufi} A. {et~al.}, 2013, \aap, 560, A105

\bibitem[{Glickman \& van(2007)}]{Glickman2007}
Glickman M., van D.~D., 2007, Methods Mol Biol, 404, 319

\bibitem[{{Guenther} {et~al}\mbox{.}(2007){Guenther}, {Esposito}, {Mundt},
  {Covino}, {Alcal{\'a}}, {Cusano}, \& {Stecklum}}]{Guenther2007}
{Guenther} E.~W., {Esposito} M., {Mundt} R., {Covino} E., {Alcal{\'a}} J.~M.,
  {Cusano} F., {Stecklum} B., 2007, \aap, 467, 1147

\bibitem[{{Harris}, {Morgan} \& {Roman}(1954){Harris}, {Morgan}, \&
  {Roman}}]{Harris1954}
{Harris} D.~L., {Morgan} W.~W., {Roman} N.~G., 1954, \apj, 119, 622

\bibitem[{{Harris} {et~al}\mbox{.}(2012){Harris}, {Andrews}, {Wilner}, \&
  {Kraus}}]{Harris2012}
{Harris} R.~J., {Andrews} S.~M., {Wilner} D.~J., {Kraus} A.~L., 2012, \apj,
  751, 115

\bibitem[{{Hartmann}(2001)}]{Hartmann2001}
{Hartmann} L., 2001, \aj, 121, 1030

\bibitem[{{Herbig} \& {Jones}(1983)}]{Herbig1983}
{Herbig} G.~H., {Jones} B.~F., 1983, \aj, 88, 1040

\bibitem[{{Herczeg} \& {Hillenbrand}(2008)}]{Herczeg2008}
{Herczeg} G.~J., {Hillenbrand} L.~A., 2008, \apj, 681, 594

\bibitem[{{Hirota}(2010)}]{Hirota2010}
{Hirota} T., 2010, in 10th European VLBI Network Symposium and EVN Users
  Meeting: VLBI and the New Generation of Radio Arrays, p.~63

\bibitem[{{Hirota} {et~al}\mbox{.}(2008){Hirota}, {Bushimata}, {Choi}, {Honma},
  {Imai}, {Iwadate}, {Jike}, {Kameya}, {Kamohara}, {Kan-Ya}, {Kawaguchi},
  {Kijima}, {Kobayashi}, {Kuji}, {Kurayama}, {Manabe}, {Miyaji}, {Nagayama},
  {Nakagawa}, {Oh}, {Omodaka}, {Oyama}, {Sakai}, {Sasao}, {Sato}, {Shibata},
  {Tamura}, \& {Yamashita}}]{Hirota2008}
{Hirota} T. {et~al.}, 2008, \pasj, 60, 37

\bibitem[{{Hu{\'e}lamo} {et~al}\mbox{.}(2015){Hu{\'e}lamo}, {de
  Gregorio-Monsalvo}, {Macias}, {Pinte}, {Ireland}, {Tuthill}, \&
  {Lacour}}]{Huelamo2015}
{Hu{\'e}lamo} N., {de Gregorio-Monsalvo} I., {Macias} E., {Pinte} C., {Ireland}
  M., {Tuthill} P., {Lacour} S., 2015, \aap, 575, L5

\bibitem[{{Hu{\'e}lamo} {et~al}\mbox{.}(2011){Hu{\'e}lamo}, {Lacour},
  {Tuthill}, {Ireland}, {Kraus}, \& {Chauvin}}]{Huelamo2011}
{Hu{\'e}lamo} N., {Lacour} S., {Tuthill} P., {Ireland} M., {Kraus} A.,
  {Chauvin} G., 2011, \aap, 528, L7

\bibitem[{{Ireland} {et~al}\mbox{.}(2008){Ireland}, {Kraus}, {Martinache},
  {Lloyd}, \& {Tuthill}}]{Ireland2008a}
{Ireland} M.~J., {Kraus} A., {Martinache} F., {Lloyd} J.~P., {Tuthill} P.~G.,
  2008, \apj, 678, 463

\bibitem[{{Ireland} \& {Kraus}(2008)}]{Ireland2008}
{Ireland} M.~J., {Kraus} A.~L., 2008, \apjl, 678, L59

\bibitem[{{Isella} \& {Natta}(2005)}]{Isella2005}
{Isella} A., {Natta} A., 2005, \aap, 438, 899

\bibitem[{{Janson} {et~al}\mbox{.}(2012){Janson}, {Hormuth}, {Bergfors},
  {Brandner}, {Hippler}, {Daemgen}, {Kudryavtseva}, {Schmalzl}, {Schnupp}, \&
  {Henning}}]{Janson2012}
{Janson} M. {et~al.}, 2012, \apj, 754, 44

\bibitem[{Jeffreys(1998)}]{Jeffreys1961}
Jeffreys H., 1998, Theory of probability, Oxford Classic Texts in the Physical
  Sciences. The Clarendon Press, Oxford University Press, New York, pp.
  xii+459, reprint of the 1983 edition

\bibitem[{{Kenyon} \& {Hartmann}(1995)}]{Kenyon1995}
{Kenyon} S.~J., {Hartmann} L., 1995, \apjs, 101, 117

\bibitem[{{Kley}(2000)}]{Kley2000}
{Kley} W., 2000, in IAU Symposium, Vol. 200, IAU Symposium, p. 211

\bibitem[{{Kley} \& {Nelson}(2007)}]{Kley2007}
{Kley} W., {Nelson} R., 2007, ArXiv e-prints

\bibitem[{{Koepferl} {et~al}\mbox{.}(2013){Koepferl}, {Ercolano}, {Dale},
  {Teixeira}, {Ratzka}, \& {Spezzi}}]{Koepferl2013}
{Koepferl} C.~M., {Ercolano} B., {Dale} J., {Teixeira} P.~S., {Ratzka} T.,
  {Spezzi} L., 2013, \mnras, 428, 3327

\bibitem[{{Kohn} {et~al}\mbox{.}(2016){Kohn}, {Shkolnik}, {Weinberger},
  {Carlberg}, \& {Llama}}]{Kohn2016}
{Kohn} S.~A., {Shkolnik} E.~L., {Weinberger} A.~J., {Carlberg} J.~K., {Llama}
  J., 2016, \apj, 820, 2

\bibitem[{{Kraus} {et~al}\mbox{.}(2015){Kraus}, {Andrews}, {Bowler}, {Herczeg},
  {Ireland}, {Liu}, {Metchev}, \& {Cruz}}]{Kraus2015_FWTau}
{Kraus} A.~L., {Andrews} S.~M., {Bowler} B.~P., {Herczeg} G., {Ireland} M.~J.,
  {Liu} M.~C., {Metchev} S., {Cruz} K.~L., 2015, \apjl, 798, L23

\bibitem[{{Kraus} {et~al}\mbox{.}(2012){Kraus}, {Ireland}, {Hillenbrand}, \&
  {Martinache}}]{Kraus2012}
{Kraus} A.~L., {Ireland} M.~J., {Hillenbrand} L.~A., {Martinache} F., 2012,
  \apj, 745, 19

\bibitem[{{Kraus} {et~al}\mbox{.}(2016){Kraus}, {Ireland}, {Huber}, {Mann}, \&
  {Dupuy}}]{Kraus2016}
{Kraus} A.~L., {Ireland} M.~J., {Huber} D., {Mann} A.~W., {Dupuy} T.~J., 2016,
  ArXiv e-prints

\bibitem[{{Lada}, {Lombardi} \& {Alves}(2009){Lada}, {Lombardi}, \&
  {Alves}}]{Lada2009}
{Lada} C.~J., {Lombardi} M., {Alves} J.~F., 2009, \apj, 703, 52

\bibitem[{{Le Blanc}, {Covey} \& {Stassun}(2011){Le Blanc}, {Covey}, \&
  {Stassun}}]{LeBlanc2011}
{Le Blanc} T.~S., {Covey} K.~R., {Stassun} K.~G., 2011, \aj, 142, 55

\bibitem[{{Liu} {et~al}\mbox{.}(2015){Liu}, {Joergens}, {Bayo}, {Nielbock}, \&
  {Wang}}]{Liu2015}
{Liu} Y., {Joergens} V., {Bayo} A., {Nielbock} M., {Wang} H., 2015, \aap, 582,
  A22

\bibitem[{{Loinard} {et~al}\mbox{.}(2008){Loinard}, {Torres}, {Mioduszewski},
  \& {Rodr{\'{\i}}guez}}]{Loinard2008}
{Loinard} L., {Torres} R.~M., {Mioduszewski} A.~J., {Rodr{\'{\i}}guez} L.~F.,
  2008, in IAU Symposium, Vol. 248, IAU Symposium, {Jin} W.~J., {Platais} I.,
  {Perryman} M.~A.~C., eds., pp. 186--189

\bibitem[{{Lubow} \& {Artymowicz}(2000)}]{Lubow2000}
{Lubow} S.~H., {Artymowicz} P., 2000, Protostars and Planets IV, 731

\bibitem[{{Lubow}, {Seibert} \& {Artymowicz}(1999){Lubow}, {Seibert}, \&
  {Artymowicz}}]{Lubow1999}
{Lubow} S.~H., {Seibert} M., {Artymowicz} P., 1999, \apj, 526, 1001

\bibitem[{{Luhman} {et~al}\mbox{.}(2009){Luhman}, {Mamajek}, {Allen}, \&
  {Cruz}}]{Luhman2009}
{Luhman} K.~L., {Mamajek} E.~E., {Allen} P.~R., {Cruz} K.~L., 2009, \apj, 703,
  399

\bibitem[{{Luhman} {et~al}\mbox{.}(1998){Luhman}, {Rieke}, {Lada}, \&
  {Lada}}]{Luhman1998}
{Luhman} K.~L., {Rieke} G.~H., {Lada} C.~J., {Lada} E.~A., 1998, \apj, 508, 347

\bibitem[{{Luhman} {et~al}\mbox{.}(2003){Luhman}, {Stauffer}, {Muench},
  {Rieke}, {Lada}, {Bouvier}, \& {Lada}}]{Luhman2003}
{Luhman} K.~L., {Stauffer} J.~R., {Muench} A.~A., {Rieke} G.~H., {Lada} E.~A.,
  {Bouvier} J., {Lada} C.~J., 2003, \apj, 593, 1093

\bibitem[{{Mamajek}(2008)}]{Mamajek2008}
{Mamajek} E.~E., 2008, Astronomische Nachrichten, 329, 10

\bibitem[{{Martinache}(2011)}]{Martinache2011}
{Martinache} F., 2011, in Society of Photo-Optical Instrumentation Engineers
  (SPIE) Conference Series, Vol. 8151, Society of Photo-Optical Instrumentation
  Engineers (SPIE) Conference Series

\bibitem[{{Mathis}(1990)}]{Mathis1990}
{Mathis} J.~S., 1990, in Astronomical Society of the Pacific Conference Series,
  Vol.~12, The Evolution of the Interstellar Medium, {Blitz} L., ed., pp.
  63--77

\bibitem[{{Meeus}(1992)}]{Meeus1992}
{Meeus} J., 1992, Journal of the British Astronomical Association, 102, 109

\bibitem[{{Metchev} \& {Hillenbrand}(2009)}]{Metchev2009}
{Metchev} S.~A., {Hillenbrand} L.~A., 2009, \apjs, 181, 62

\bibitem[{{Muzerolle} {et~al}\mbox{.}(2010){Muzerolle}, {Allen}, {Megeath},
  {Hern{\'a}ndez}, \& {Gutermuth}}]{Muzerolle2010}
{Muzerolle} J., {Allen} L.~E., {Megeath} S.~T., {Hern{\'a}ndez} J., {Gutermuth}
  R.~A., 2010, \apj, 708, 1107

\bibitem[{{Najita}, {Andrews} \& {Muzerolle}(2015){Najita}, {Andrews}, \&
  {Muzerolle}}]{Najita2015}
{Najita} J.~R., {Andrews} S.~M., {Muzerolle} J., 2015, \mnras, 450, 3559

\bibitem[{{Nomura} {et~al}\mbox{.}(2016){Nomura}, {Tsukagoshi}, {Kawabe},
  {Ishimoto}, {Okuzumi}, {Muto}, {Kanagawa}, {Ida}, {Walsh}, {Millar}, \&
  {Bai}}]{Nomura2016}
{Nomura} H. {et~al.}, 2016, \apjl, 819, L7

\bibitem[{{Olofsson} {et~al}\mbox{.}(2013){Olofsson}, {Sz{\H u}cs}, {Henning},
  {Linz}, {Pascucci}, \& {Joergens}}]{Olofsson2013}
{Olofsson} J., {Sz{\H u}cs} L., {Henning} T., {Linz} H., {Pascucci} I.,
  {Joergens} V., 2013, \aap, 560, A100

\bibitem[{{Orellana} {et~al}\mbox{.}(2012){Orellana}, {Cieza}, {Schreiber},
  {Mer{\'{\i}}n}, {Brown}, {Pellizza}, \& {Romero}}]{Orellana2012}
{Orellana} M., {Cieza} L.~A., {Schreiber} M.~R., {Mer{\'{\i}}n} B., {Brown}
  J.~M., {Pellizza} L.~J., {Romero} G.~A., 2012, \aap, 539, A41

\bibitem[{{Pascucci} {et~al}\mbox{.}(2008){Pascucci}, {Apai},
  {Hardegree-Ullman}, {Kim}, {Meyer}, \& {Bouwman}}]{Pascucci2008}
{Pascucci} I., {Apai} D., {Hardegree-Ullman} E.~E., {Kim} J.~S., {Meyer} M.~R.,
  {Bouwman} J., 2008, \apj, 673, 477

\bibitem[{{Pecaut} \& {Mamajek}(2013)}]{Pecaut2013}
{Pecaut} M.~J., {Mamajek} E.~E., 2013, \apjs, 208, 9

\bibitem[{{Pinilla}, {Benisty} \& {Birnstiel}(2012){Pinilla}, {Benisty}, \&
  {Birnstiel}}]{Pinilla2012}
{Pinilla} P., {Benisty} M., {Birnstiel} T., 2012, \aap, 545, A81

\bibitem[{{Pollack} {et~al}\mbox{.}(1996){Pollack}, {Hubickyj}, {Bodenheimer},
  {Lissauer}, {Podolak}, \& {Greenzweig}}]{Pollack1996}
{Pollack} J.~B., {Hubickyj} O., {Bodenheimer} P., {Lissauer} J.~J., {Podolak}
  M., {Greenzweig} Y., 1996, \icarus, 124, 62

\bibitem[{{Pott} {et~al}\mbox{.}(2010){Pott}, {Perrin}, {Furlan}, {Ghez},
  {Herbst}, \& {Metchev}}]{Pott2010}
{Pott} J.-U., {Perrin} M.~D., {Furlan} E., {Ghez} A.~M., {Herbst} T.~M.,
  {Metchev} S., 2010, \apj, 710, 265

\bibitem[{{Pravdo} {et~al}\mbox{.}(2006){Pravdo}, {Shaklan}, {Wiktorowicz},
  {Kulkarni}, {Lloyd}, {Martinache}, {Tuthill}, \& {Ireland}}]{Pravdo2006}
{Pravdo} S.~H., {Shaklan} S.~B., {Wiktorowicz} S.~J., {Kulkarni} S., {Lloyd}
  J.~P., {Martinache} F., {Tuthill} P.~G., {Ireland} M.~J., 2006, \apj, 649,
  389

\bibitem[{{Raghavan} {et~al}\mbox{.}(2010){Raghavan}, {McAlister}, {Henry},
  {Latham}, {Marcy}, {Mason}, {Gies}, {White}, \& {ten
  Brummelaar}}]{Raghavan2010}
{Raghavan} D. {et~al.}, 2010, \apjs, 190, 1

\bibitem[{{Rapson} {et~al}\mbox{.}(2015){Rapson}, {Kastner},
  {Millar-Blanchaer}, \& {Dong}}]{Rapson2015}
{Rapson} V.~A., {Kastner} J.~H., {Millar-Blanchaer} M.~A., {Dong} R., 2015,
  ArXiv e-prints

\bibitem[{{Rebollido} {et~al}\mbox{.}(2015){Rebollido}, {Mer{\'{\i}}n},
  {Ribas}, {Bustamante}, {Bouy}, {Riviere-Marichalar}, {Prusti}, {Pilbratt},
  {Andr{\'e}}, \& {{\'A}brah{\'a}m}}]{Rebollido2015}
{Rebollido} I. {et~al.}, 2015, ArXiv e-prints

\bibitem[{{Rebull} {et~al}\mbox{.}(2010){Rebull}, {Padgett}, {McCabe},
  {Hillenbrand}, {Stapelfeldt}, {Noriega-Crespo}, {Carey}, {Brooke}, {Huard},
  {Terebey}, {Audard}, {Monin}, {Fukagawa}, {G{\"u}del}, {Knapp}, {Menard},
  {Allen}, {Angione}, {Baldovin-Saavedra}, {Bouvier}, {Briggs}, {Dougados},
  {Evans}, {Flagey}, {Guieu}, {Grosso}, {Glauser}, {Harvey}, {Hines}, {Latter},
  {Skinner}, {Strom}, {Tromp}, \& {Wolf}}]{Rebull2010}
{Rebull} L.~M. {et~al.}, 2010, \apjs, 186, 259

\bibitem[{{Rice} {et~al}\mbox{.}(2006{\natexlab{a}}){Rice}, {Armitage}, {Wood},
  \& {Lodato}}]{Rice2006a}
{Rice} W.~K.~M., {Armitage} P.~J., {Wood} K., {Lodato} G., 2006{\natexlab{a}},
  \mnras, 373, 1619

\bibitem[{{Rice} {et~al}\mbox{.}(2006{\natexlab{b}}){Rice}, {Lodato},
  {Pringle}, {Armitage}, \& {Bonnell}}]{Rice2006}
{Rice} W.~K.~M., {Lodato} G., {Pringle} J.~E., {Armitage} P.~J., {Bonnell}
  I.~A., 2006{\natexlab{b}}, \mnras, 372, L9

\bibitem[{{Rosotti}, {Ercolano} \& {Owen}(2015){Rosotti}, {Ercolano}, \&
  {Owen}}]{Rosotti2015}
{Rosotti} G.~P., {Ercolano} B., {Owen} J.~E., 2015, ArXiv e-prints

\bibitem[{{Scholz} {et~al}\mbox{.}(1999){Scholz}, {Brunzendorf}, {Ivanov},
  {Kharchenko}, {Lasker}, {Meusinger}, {Preibisch}, {Schilbach}, \&
  {Zinnecker}}]{Scholz1999}
{Scholz} R.-D. {et~al.}, 1999, \aaps, 137, 305

\bibitem[{{Sicilia-Aguilar} {et~al}\mbox{.}(2006){Sicilia-Aguilar}, {Hartmann},
  {Calvet}, {Megeath}, {Muzerolle}, {Allen}, {D'Alessio}, {Mer{\'{\i}}n},
  {Stauffer}, {Young}, \& {Lada}}]{SiciliaAguilar2006}
{Sicilia-Aguilar} A. {et~al.}, 2006, \apj, 638, 897

\bibitem[{{Siess}(2001)}]{Siess2001}
{Siess} L., 2001, in Astronomical Society of the Pacific Conference Series,
  Vol. 243, From Darkness to Light: Origin and Evolution of Young Stellar
  Clusters, {Montmerle} T., {Andr{\'e}} P., eds., p. 581

\bibitem[{{Simon} {et~al}\mbox{.}(1995){Simon}, {Ghez}, {Leinert}, {Cassar},
  {Chen}, {Howell}, {Jameson}, {Matthews}, {Neugebauer}, \&
  {Richichi}}]{Simon1995}
{Simon} M. {et~al.}, 1995, \apj, 443, 625

\bibitem[{{Skrutskie} {et~al}\mbox{.}(2006){Skrutskie}, {Cutri}, {Stiening},
  {Weinberg}, {Schneider}, {Carpenter}, {Beichman}, {Capps}, {Chester},
  {Elias}, {Huchra}, {Liebert}, {Lonsdale}, {Monet}, {Price}, {Seitzer},
  {Jarrett}, {Kirkpatrick}, {Gizis}, {Howard}, {Evans}, {Fowler}, {Fullmer},
  {Hurt}, {Light}, {Kopan}, {Marsh}, {McCallon}, {Tam}, {Van Dyk}, \&
  {Wheelock}}]{Skrutskie2006}
{Skrutskie} M.~F. {et~al.}, 2006, \aj, 131, 1163

\bibitem[{{Strom} {et~al}\mbox{.}(1989){Strom}, {Strom}, {Edwards}, {Cabrit},
  \& {Skrutskie}}]{Strom1989}
{Strom} K.~M., {Strom} S.~E., {Edwards} S., {Cabrit} S., {Skrutskie} M.~F.,
  1989, \aj, 97, 1451

\bibitem[{{Sun} {et~al}\mbox{.}(2006){Sun}, {Kramer}, {Ossenkopf}, {Bensch},
  {Stutzki}, \& {Miller}}]{Sun2006}
{Sun} K., {Kramer} C., {Ossenkopf} V., {Bensch} F., {Stutzki} J., {Miller} M.,
  2006, \aap, 451, 539

\bibitem[{{Vacca}, {Sheehy} \& {Graham}(2007){Vacca}, {Sheehy}, \&
  {Graham}}]{Vacca2007}
{Vacca} W.~D., {Sheehy} C.~D., {Graham} J.~R., 2007, \apj, 662, 272

\bibitem[{{van der Marel} {et~al}\mbox{.}(2016){van der Marel}, {van Dishoeck},
  {Bruderer}, {Andrews}, {Pontoppidan}, {Herczeg}, {van Kempen}, \&
  {Miotello}}]{vanderMarel2016}
{van der Marel} N., {van Dishoeck} E.~F., {Bruderer} S., {Andrews} S.~M.,
  {Pontoppidan} K.~M., {Herczeg} G.~J., {van Kempen} T., {Miotello} A., 2016,
  \aap, 585, A58

\bibitem[{{van der Marel} {et~al}\mbox{.}(2015){van der Marel}, {van Dishoeck},
  {Bruderer}, {P{\'e}rez}, \& {Isella}}]{VanderMarel2015}
{van der Marel} N., {van Dishoeck} E.~F., {Bruderer} S., {P{\'e}rez} L.,
  {Isella} A., 2015, \aap, 579, A106

\bibitem[{{Wilking}, {Gagn{\'e}} \& {Allen}(2008){Wilking}, {Gagn{\'e}}, \&
  {Allen}}]{Wilking2008}
{Wilking} B.~A., {Gagn{\'e}} M., {Allen} L.~E., 2008, {Star Formation in the
  {$\rho$} Ophiuchi Molecular Cloud}, {Reipurth} B., ed., p. 351

\bibitem[{{Yelda} {et~al}\mbox{.}(2010){Yelda}, {Lu}, {Ghez}, {Clarkson},
  {Anderson}, {Do}, \& {Matthews}}]{Yelda2010}
{Yelda} S., {Lu} J.~R., {Ghez} A.~M., {Clarkson} W., {Anderson} J., {Do} T.,
  {Matthews} K., 2010, \apj, 725, 331

\end{thebibliography}

\end{document}